\documentclass[useAMS,usenatbib,letterpaper,10pt]{mn2e} 
\newcommand{\hmpc}{{\,h^{-1}\rm\,Mpc\,}}

\newcommand{\msun}{\rm{M}_{\odot}}

\newcommand{\mstars}{{\rm M}_{\rm stellar}}
\newcommand{\mhalo}{{\rm M}_{\rm halo}}

\usepackage{graphicx}
\usepackage{natbib}
\usepackage{aas_macros}
\usepackage{amsmath}
\usepackage{amssymb}
\usepackage{fancyvrb}
\usepackage{epsf}
\usepackage{sidecap}
\usepackage{color}
\usepackage{times}
\usepackage{epsfig}
\usepackage[T1]{fontenc}
\usepackage{aecompl}
\usepackage{comment}
\usepackage[bookmarks,bookmarksnumbered,colorlinks=true, citecolor=blue, linkcolor=black]{hyperref}
\title[Galactic Conformity in Illustris]{Modeling Galactic Conformity with the Color--Halo Age Relation in the Illustris Simulation}

\author[Bray et al.]
{Aaron~D.~Bray$^{1}$\thanks{E-mail:abray@cfa.harvard.edu}, 
Annalisa Pillepich$^{1}$,
Laura~V.~Sales$^{2}$,
Emily~Zhu$^{3,4}$,
Shy~Genel$^{5}$\thanks{Hubble Fellow},
\newauthor Vicente~Rodriguez-Gomez$^{1}$,
Paul~Torrey$^{6,7}$,
Dylan~Nelson$^{1}$,
Mark~Vogelsberger$^{6}$,
\newauthor 
Volker~Springel$^{8,9}$,
Daniel~J.~Eisenstein$^{1}$,
and Lars~Hernquist$^{1}$\vspace{2mm}
\\
$^{1}$Harvard-Smithsonian Center for Astrophysics, 60 Garden Street, Cambridge, MA 02138, USA\\
$^{2}$University of California Riverside, 900 University Avenue, Riverside, CA 92521, USA\\
$^{3}$Phillips Academy, 180 Main Street, Andover, MA 01810, USA\\
$^{4}$Harvard College, Cambridge, MA 02138\\
$^{5}$Columbia University, 116th Street \& Broadway, New York, NY 10027, USA\\
$^{6}$Massachusetts Institute of Technology, 77 Massachusetts Avenue, Cambridge, MA 02139, USA\\
$^{7}$TAPIR, Mailcode 350-17, California Institute of Technology, Pasadena, CA 91125, USA\\
$^{8}$Heidelberg Institute for Theoretical Studies, Schloss-Wolfsbrunnenweg 35, 69118 Heidelberg, Germany\\
$^{9}$Zentrum fur Astronomie der Universitat Heidelberg, ARI, Monchhofstr. 12-14, 69120 Heidelberg, Germany}

\begin{document}

\pagerange{\pageref{firstpage}--\pageref{lastpage}} \pubyear{2015}

\maketitle

\label{firstpage}


\begin{abstract}

Comparisons between observational surveys and galaxy formation models find that dark matter haloes' mass can largely explain their galaxies' stellar mass. However, it remains uncertain whether additional environmental variables, known as assembly bias, are necessary to explain other galaxy properties. We use the Illustris Simulation to investigate the role of assembly bias in producing galactic conformity by considering 18,000 galaxies with $\mstars  > 2 \times 10^9 \msun$. We find a significant signal of galactic conformity: out to distances of about 10 Mpc, the mean red fraction of galaxies around redder galaxies is higher than around bluer galaxies $\textit{at fixed stellar mass}$. Dark matter haloes exhibit an analogous conformity signal, in which the fraction of haloes formed at earlier times (old haloes) is higher around old haloes than around younger ones $\textit{at fixed halo mass}$. A plausible interpretation of galactic conformity is the combination of the halo conformity signal with the galaxy color--halo age relation: at fixed stellar mass, particularly toward the low-mass end, Illustris' galaxy colors correlate with halo age, with the reddest galaxies (often satellites) preferentially found in the oldest haloes. We explain the galactic conformity effect with a simple semi-empirical model, assigning stellar mass via halo mass (abundance matching) and galaxy color via halo age (age matching). Regarding comparison to observations, we conclude that the adopted selection/isolation criteria, projection effects, and stacking techniques can have a significant impact on the measured amplitude of the conformity signal.

\end{abstract}

\begin{keywords}
galaxies: formation -- galaxies: clustering -- galaxies: haloes -- cosmology: dark matter -- cosmology: theory -- cosmology: simulations
\end{keywords}


\section{Introduction}
\label{sec:intro}

Previous investigations of the demographics and distribution of dark matter haloes in a cold dark matter universe have found that the clustering properties of these haloes have a dependence on formation time, in addition to the more significant dependence on halo mass \citep{Gao05, Wechsler06, Croton07, Li08}. However, current observational frameworks for analyzing the luminosity- and color-dependent clustering of galaxies do not take into account this halo assembly bias (e.g., \citealt{Zehavi11}). Rather, they use models that assume that galaxy clustering statistics can be modeled solely based on the mass of the halo (e.g., \citealt{BW02}; \citealt{Yang03}; \citealt{Conroy06}). This would be a correct assumption, as long as galaxy properties such as stellar mass and specific star formation rate (sSFR) are not also correlated with other dark matter properties at fixed halo mass. Otherwise, ignoring the effects of properties other than mass may lead to biased interpretations of the observational results \citep{Zentner14}.

Recent observations at low redshift have found a signal of \textit{galactic conformity} in which the sSFR and gas fractions of neighboring galaxies correlate with the respective properties of the central galaxy, both within and beyond the virial radius (e.g., \citealt{Weinmann06}; \citealt{Kauffmann13}; \citealt{Lacerna14}; \citealt{Hartley15}; \citealt{Knobel15}). Such observations suggest that those galaxy properties may indeed be correlated with halo properties beyond mass, such that \textit{halo} assembly bias may lead to \textit{galaxy} assembly bias. Simultaneously, though, other observations have not found these same manifestations of galaxy assembly bias (e.g., \citealt{Tinker11}; \citealt{Lin15}). Moreover, there is considerable debate as to what the role of central and satellite galaxies play in the emergence of this signal \citep{Knobel15}. Some models treat centrals and satellites identically \citep{Hearin14}, while others have satellites colors correlate directly with group-wide properties, such as halo concentration \citep{Paranjape15}.  Finally, the debate over the role of internal (e.g., \citealt{Hartley15}) versus external (e.g., \citealt{Hearin15}) quenching mechanisms, and thus also the extent to which conformity is a product of assembly bias, relies heavily on the observed amplitude and radius out to which the conformity signal is observed (\citealt{Knobel15}; \citealt{Paranjape15}).

Semi-analytic models \citep{Guo11} can qualitatively produce the galactic conformity effect seen in observations, but it has been argued that such theoretical effects are not as large as in observations (\citealt{Kauffmann13}; \citealt{Hearin15}). The same qualitative conformity signal can also be reproduced using semi-empirical halo occupation models (\citealt{HW13}; \citealt{Watson14}; \citealt{Hearin15}) or with tunable extensions to the Halo Occupation Distribution framework \citep[HOD, ][]{Paranjape15}.

In this paper, we investigate the presence of galactic conformity in Illustris \citep{Vogelsberger14a}, a state-of-the-art cosmological simulation with full hydrodynamical and sub-grid physics run with the {\sc arepo} code \citep{Springel10}. Galactic conformity has yet to be probed in a hydrodynamical simulation, given the limitations thus far in encompassed volumes, numerical resolution, and realism and statistical significance of the simulated galaxy populations. Illustris, on the other hand, combines a $75 \hmpc$ per side cosmological volume at kpc resolution with a population of thousands of galaxies which compare well to observational constraints. By studying galaxy clustering in Illustris, we see whether a statistically significant galactic conformity signal arises in a realistic simulation of galaxy formation, and in particular, whether the conformity can be explained solely by differences in the halo masses of red and blue galaxies, or whether the additional information about the assembly history of the haloes is required. Furthermore, we explore the role that possible observational biases and selection criteria will have on the conformity signal.

This paper is organised as follows. In Section~\ref{sec:methods}, we briefly review the properties of the Illustris Simulation, describe our selection criterion, and explain how we calculate dark matter halo ages from the merger trees. We present the detection of both galactic and halo conformity in Section~\ref{sec:conformity}, and show the presence of a color--halo age relation in Section~\ref{sec:colorage}. In Section~\ref{sec:discussion}, we apply abundance and age matching models from the literature to the Illustris galaxies to show how galactic conformity naturally arises in Illustris, and we discuss the differential importance of centrals and satellites, as well as the effect of other observational choices on the strength and radial dependence of the conformity signal. We conclude and summarize in Section~\ref{sec:conclusion}.

\begin{figure*}
	\includegraphics[width=0.44\linewidth]{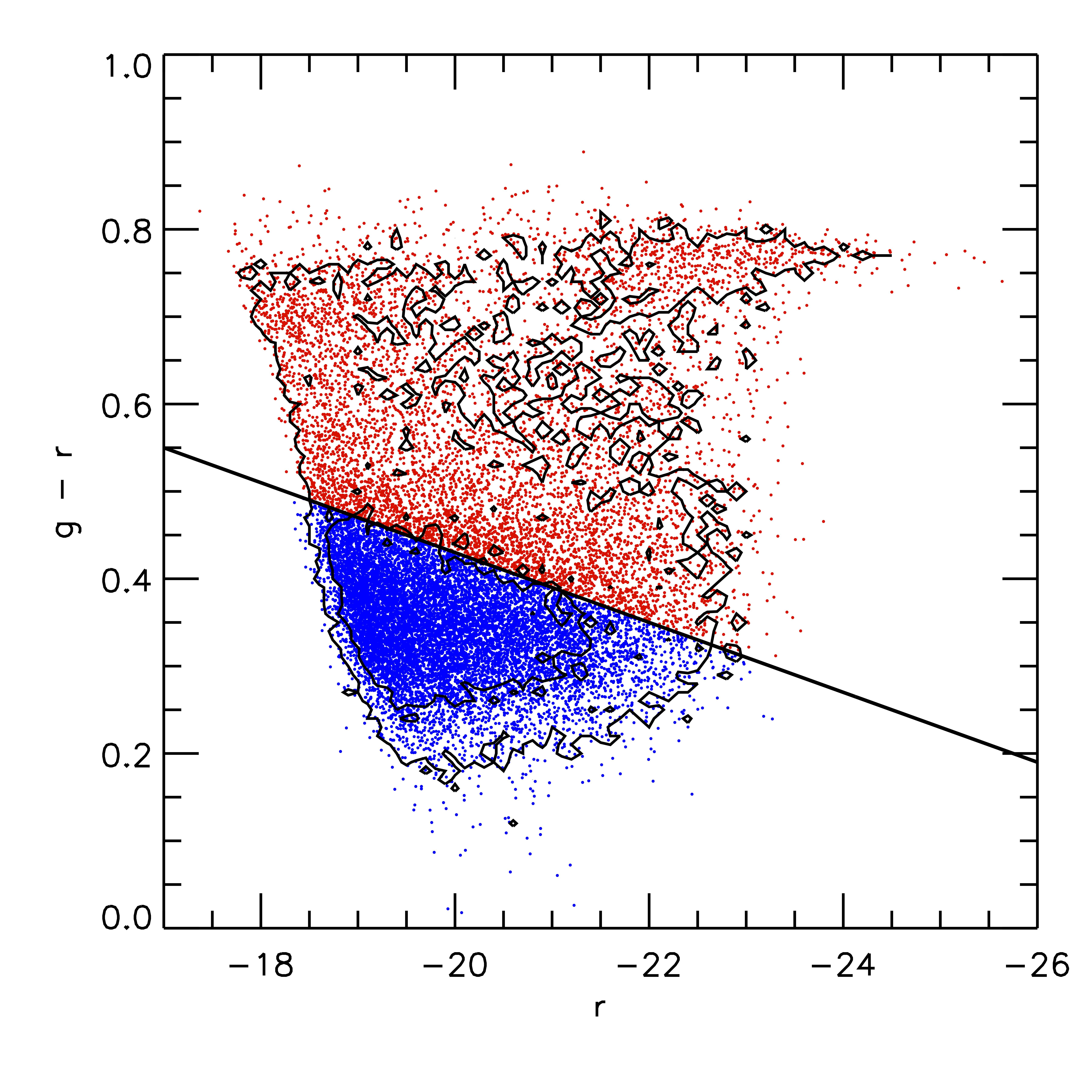}
	\includegraphics[width=0.44\linewidth]{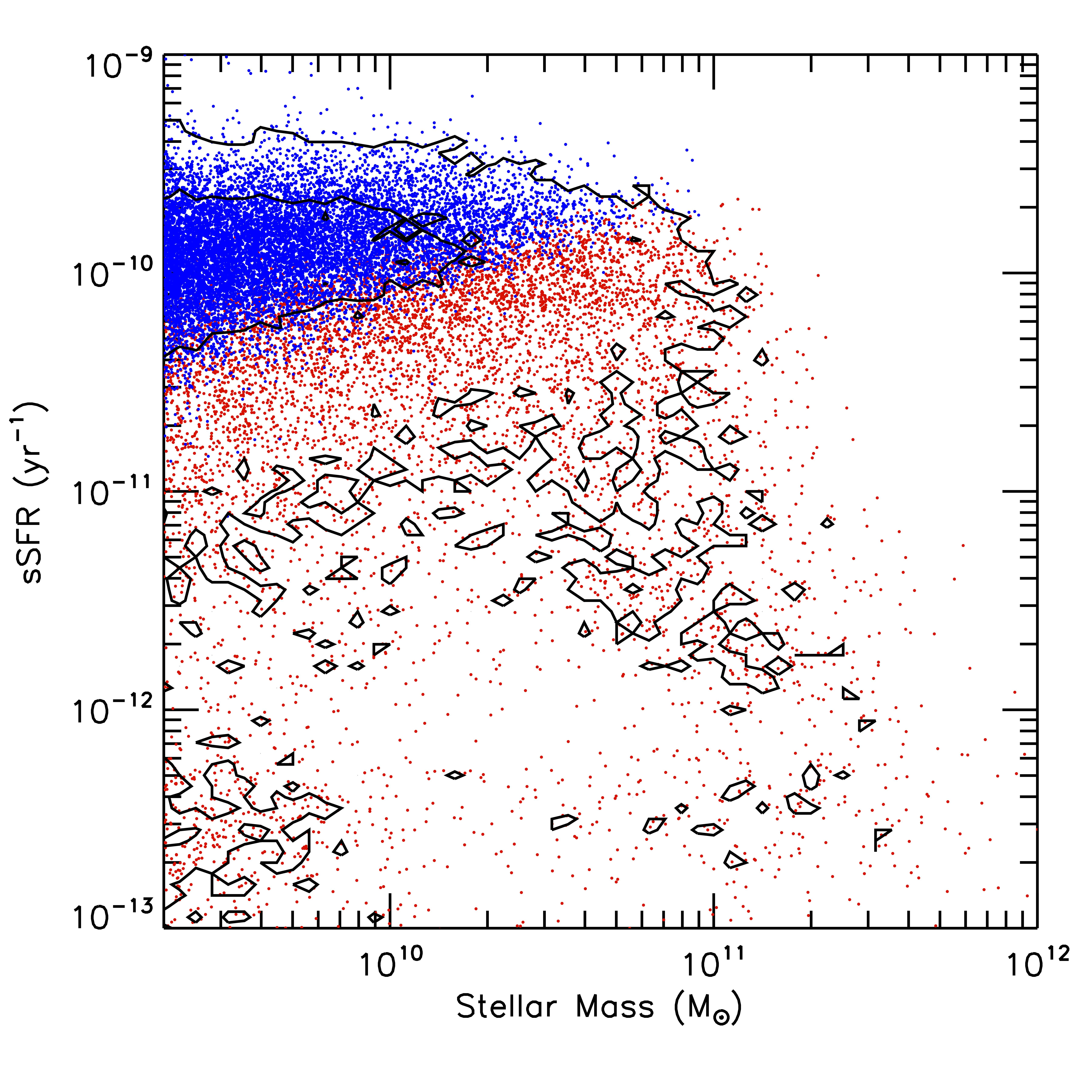}

 	\caption{The galaxy population in Illustris. Left: The color-magnitude diagram for the $\mstars > 2 \times 10^9 \msun$ galaxy sample used in this work. The solid black line divides the red and blue populations (shown in corresponding colors.) Contours for $50\%$ and $95\%$ inclusion in absolute $M_{r}$ versus $g - r$ color space are shown for our $18$, $243$ galaxies. To allow for a statistically meaningful comparison of redder and bluer galaxies, we divide the sample such that ``green valley" galaxies are grouped with red sequence galaxies, despite having some ongoing star-formation. Right: Specific star formation rates as a function of stellar mass. We show how the red-blue color-luminosity cut in Illustris translates into the sSFR-stellar mass plane.
	 While the $50\%$ and $95\%$ inclusion contours correspond to the distribution of points in the sSFR-stellar mass plane, the colors are inherited from the left plot. For clarity, we randomly plot galaxies with no star formation rate randomly distributed about $2 \times 10^{-13}$ yr$^{-1}$. 
	}
	\label{fig:cmd}
        
\end{figure*}


\section{Simulations and Methods}
\label{sec:methods}

In this paper we use the Illustris Suite, a set of simulations which form galaxies self-consistently, by combining an N-body treatment of gravity with the hydrodynamical, moving-mesh code {\sc arepo} \citep{Springel10} to follow gas. {\sc arepo} solves the Euler equations on an unstructured Voronoi tessellation, in which the mesh-generating points advect with the baryonic flow. The code includes relevant physical processes such as gas cooling \citep{Katz96}, a photoionizing background \citep{Faucher-Giguere09}, star formation \citep{SH03}, black hole seeding and feedback (\citealt{DiMatteo05}; \citealt{Springel05a}; \citealt{Sijacki07}), and chemical enrichment \citep{Wiersma09}. Full details of the applied galaxy formation and feedback model are described in \citet{Vogelsberger13b} with multi-epoch galaxy population properties being tested and presented in \citet{Torrey14}. These simulations reproduce realistic populations of galaxies, as demonstrated in previous Illustris analyses (\citealt{Vogelsberger14a}; \citealt{Vogelsberger14b}; \citealt{Genel14}). 
While the suite includes realizations with different box sizes and at different resolutions, our primary results presented here are based on the highest resolution run (Illustris-1), where a $75 \hmpc$ cosmological box is evolved from $z=127$ to $z=0$ with initial conditions consistent with WMAP-9 \citep{Hinshaw13}. The mass resolution for the dark matter is $m_{\rm DM} = 6.26 \times 10^6 M_\odot$, and for baryons it is roughly $m_{b} \sim 1.26 \times 10^6 M_\odot$. At  $z=0$, the softening lengths are roughly 1.42 kpc for dark matter particles and 0.71 kpc for stellar particles, being smaller at higher redshifts, and the hydrodynamics follows gas down to cell sizes as small as 48 pc. 

All the data from the Illustris project and associated documentation is now publicly available \citep{Nelson15}.\footnote{http://www.illustris-project.org}

\subsection{Galaxy Sample and Definitions}
\label{sec:defs}

Haloes and subhaloes in Illustris are identified using the {\sc fof} and {\sc subfind} algorithms \citep{Davis85, Springel01, Dolag09} at 136 snapshots in time.

In what follows, we work exclusively at redshift $z=0$ and select a sample of galaxies by imposing $\mstars > 2 \times 10^9 M_\odot$ (corresponding to a minimum of roughly 2000 stellar particles or 2800 stellar, dark-matter or gas elements). The sample includes both central and satellite galaxies, with satellites being {\sc subfind} subhaloes which are members of their parent {\sc fof} group regardless of their distance from the {\sc fof} center. Thus, in our parlance, central galaxies include field galaxies with no satellites of their own, and we call any {\sc subfind}-identified object a {\it halo} unless the distinction between haloes and subhaloes is relevant.
Moreover, all galaxy properties (stellar masses, star formation rates, colors) are derived from {\sc subfind}-identified stellar particles or cells within twice the stellar half-mass radius of the galaxy under consideration. Halo or total masses are defined as the peak mass of each halo's mass accretion history (see Sec~\ref{sec:mergertree} for details), including all gravitationally bound resolution elements.

In Fig. \ref{fig:cmd}, the color-magnitude diagram and the specific star formation rate distribution as a function of stellar masses are given for Illustris galaxies at the current epoch (both centrals and satellites; see also \citealt{Vogelsberger14b}; \citealt{Sparre15}). These are in qualitative agreement with observations except for the lack of a clear bimodality between red and blue galaxies, and an overpopulation of the green valley and the blue cloud with respect to the red sequence. Encouragingly, the color distribution of \textit{satellite} galaxies alone is in good agreement with observations \citep{Sales15}.

For the purposes of our conformity analysis, we divide the selected galaxies into binary \textit{red} and \textit{blue} subsamples, rather than using a continuous distribution of specific star formation rate as a proxy for conformity (see \citealt{Kauffmann13}). To the extent that the full distribution of sSFR in Illustris differs from observations, we believe such a binary division better allows us to investigate the emergence of a conformity signal in Illustris, and the effects of observational choices on the observed signal. Unless otherwise stated, our cut will be based on stellar colors, as follows:
\begin{eqnarray}
Red: (g - r)_{galaxy} & > & 0.04(r + 20) + 0.43 \\
Blue: (g - r)_{galaxy} & < & 0.04(r + 20) + 0.43
\end{eqnarray}

\begin{center}
\begin{figure*}
	\includegraphics[width=0.44\linewidth]{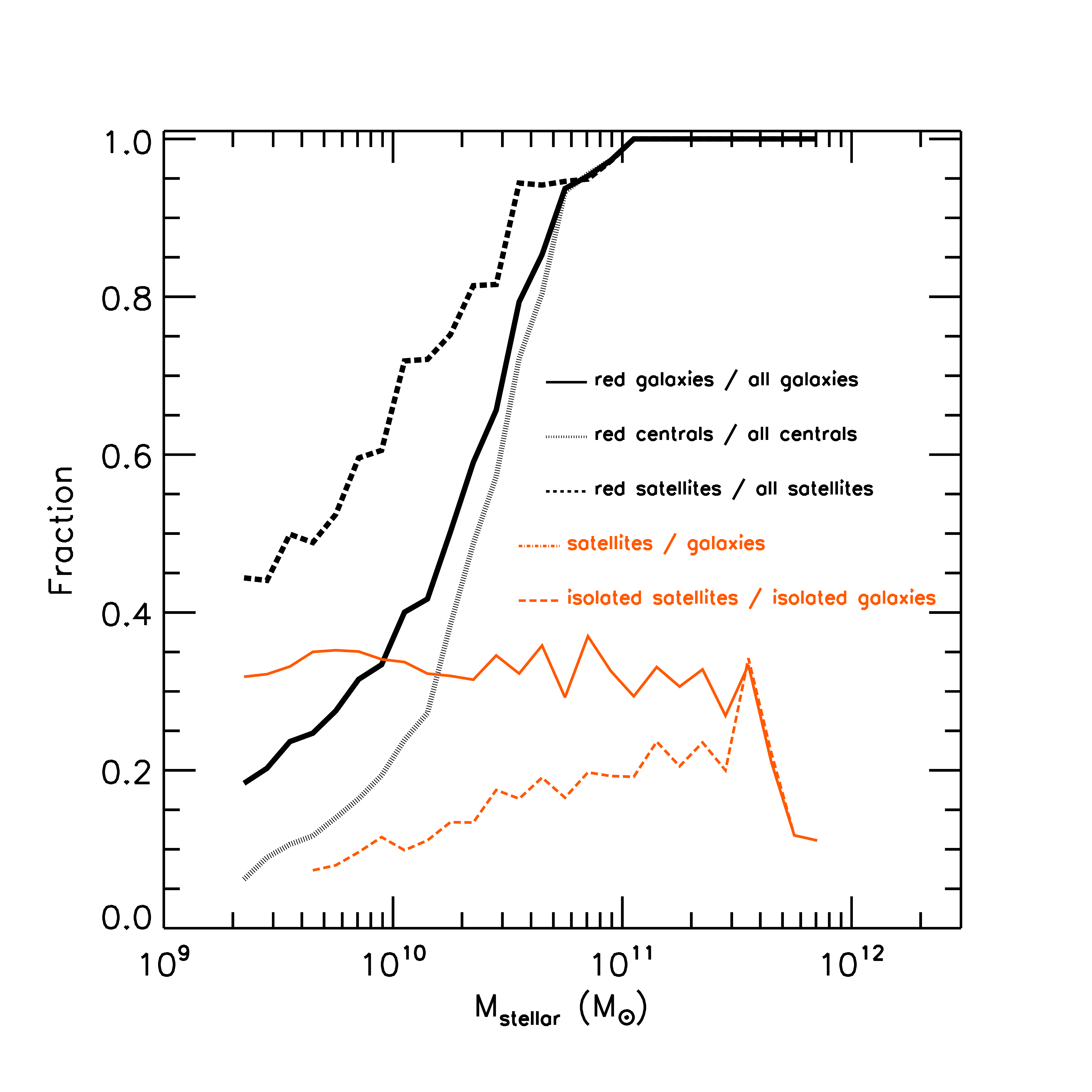}
	\includegraphics[width=0.44\linewidth]{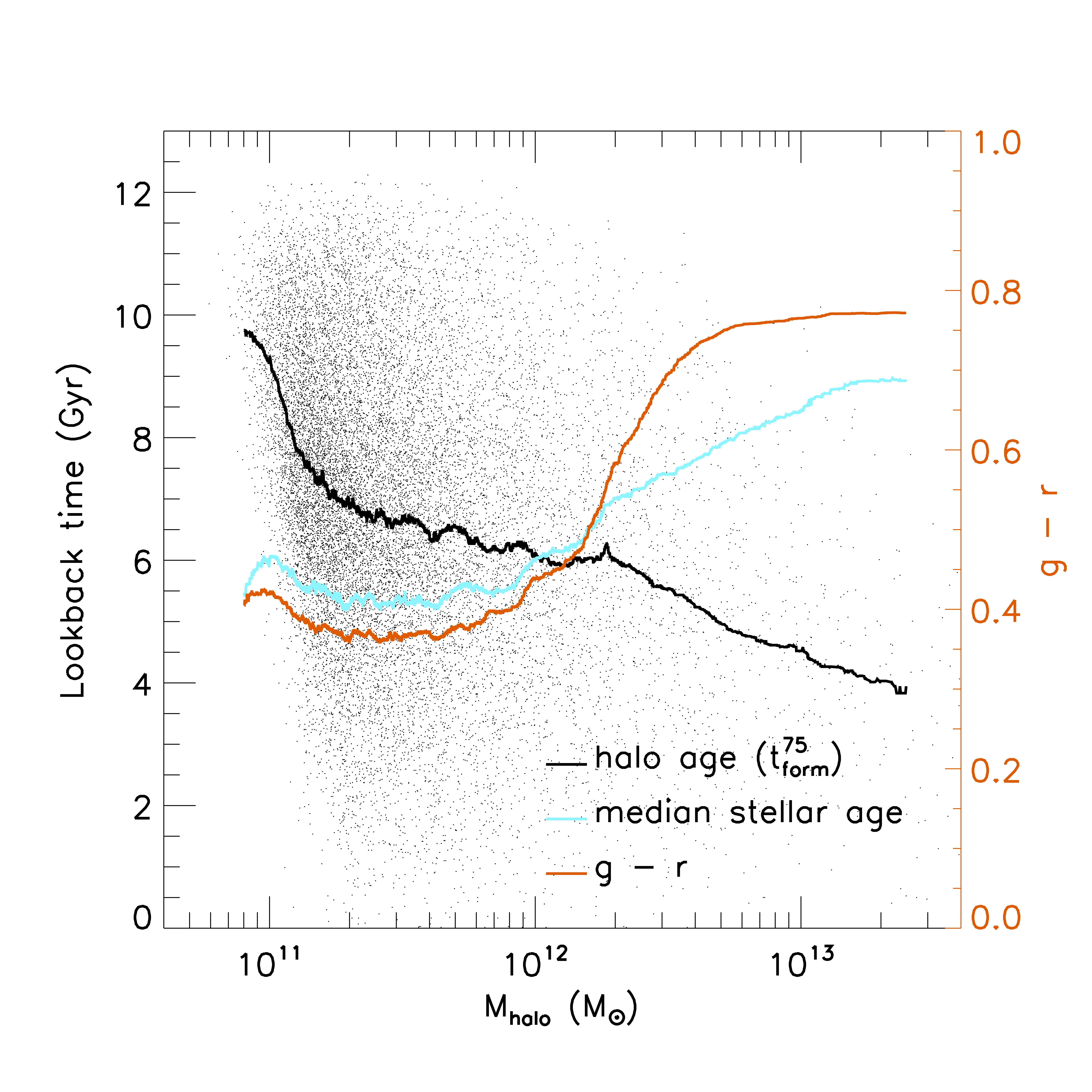}
 	
 	\caption{Left: The red fraction and satellite fraction of galaxies as a function of stellar mass. We note that the isolation criterion has the effect to reduce the satellite fraction from $\sim 30\%$ to $\sim 10\%$, although the effect is stronger at lower masses. Both centrals and satellites show a strong relationship between red fraction and halo mass; however, especially at lower masses, satellites are significantly more quenched than centrals. Right: Halo age (i.e., lookback halo formation time) as a function of halo mass, alongside both the color and median stellar age as a function of halo mass. Despite the fact that lower mass haloes are older than their more massive counterparts, they host galaxies with younger and bluer stellar populations.
	}
        
 	\label{fig:isolation_downsizing}
\end{figure*}
\end{center}

As shown in the right panel in Fig. \ref{fig:cmd}, this corresponds to a slightly increasing cut in sSFR with increasing stellar mass, with some star formation still ongoing in the red population, especially at higher masses. Moreover, in the following, we will use the terms {\it red} ({\it blue}) and {\it quenched} ({\it star-forming}) galaxies interchangeably. Finally, the fraction of red galaxies is a strong function of stellar mass, and satellite galaxies are more often red than their analog central galaxies at similar masses, in agreement with observational findings (see left panel of Fig.~\ref{fig:isolation_downsizing}).

Following observational practice (e.g., \citealt{Kauffmann13}, \citealt{Hearin14}), in the following sections, we will also adopt an isolation criterion for our sample of galaxies and divide them into {\it primaries} and {\it secondaries}. Unlike observations, we have full spatial knowledge of our galaxies, and so we define the isolation criterion based on the 3D real-space locations within the simulation volume, rather than relying on a 2D projection and redshift-space cut for the line-of-sight dimension. An Illustris galaxy is \textit{isolated} if, given its stellar mass $\mstars$, no other galaxy with stellar mass greater than $\mstars/2$ is present within a 3D distance of $500$ kpc. The \textit{primary} sample is constituted by those galaxies with $\rm{log} \, \mstars > 9.61 \approx 4.07 \times 10^9 \msun$ that are also isolated.  The adopted mass cut is the minimum for which we can consistently apply the isolation criterion, given our minimum mass threshold of $\mstars > 2 \times 10^9 \msun$. We use the term \textit{secondary} or \textit{neighboring} to refer to all galaxies in the vicinity (in our case, out to 10 Mpc) of a primary galaxy.

The purpose of the isolation criterion is to reduce the number of interloping satellite galaxies in the primary sample when the distinction between centrals and satellites is not available (as often is the case in observations). This can be seen in Fig.~\ref{fig:isolation_downsizing}, left panel, where the isolation criterion serves to lower the satellite fraction from $\sim 30\%$ down to about $10\%$, with Illustris total satellite fraction falling in between the estimates from semi-analytical models and observations \citep[][-- at higher masses, the satellite fraction drops quickly]{Kauffmann13, Guo11, Wang13}. While a secondary galaxy need not be isolated, it may be. Thus, secondary galaxies around a particular primary galaxy may themselves be members of the primary sample, since the conformity signal is measured out to radii well beyond the 500 kpc radius used for the isolation criterion.

For the conformity itself, we measure the quenched fraction of secondary galaxies in every 3D real-space radial bin R around primaries. We then report the mean value of this red fraction for primary galaxies that have at least one galaxy in radial bin R. Thus, if a particular primary has no neighboring secondary galaxy in a particular radial bin, this does not count toward the mean. 
Another way of saying this is that the mean red fraction of primary galaxies is not equivalent to the red fraction of the stack of all primary galaxies. This distinction means that conformity, as we measure it, will equally weight galaxies with few satellites and galaxies with many satellites, rather than letting the signal be dominated by a few primary galaxies with the highest halo-to-stellar mass ratios.
We report our galactic conformity results in three bins in primary stellar mass and three bins in primary halo mass, so that we can discern effects on the conformity signal caused by differences in stellar-to-halo mass relation of red versus blue galaxies.

\subsection{Halo Merger Tree and Assembly Histories}
\label{sec:mergertree}

To follow the evolution of individual haloes and galaxies, we use the {\sc SubLink} merger tree catalogs from \citet{Rodriguez-Gomez15}. These merger trees provide the evolution of any {\sc subfind} property along the main branch of all haloes and galaxies at $z = 0$. Specifically, the \textit{main branch} is defined as the sequence of progenitors with the most massive history behind them (rather than the sequence of progenitors which maximize the mass at every time step). While different definitions of main branch are on average consistent, the addition utilized here provides a safeguard against spurious defects in the halo finding algorithms, such as subhalo swapping.

We use the total {\sc subfind} mass accretion histories to calculate the halo formation time or halo age, $t_{\rm form}$, of every halo, as well as the halo mass. In practice, for every object within our sample at $z=0$, we first run the sequence of masses at subsequent snapshots through a median box filter of full width of five snapshots (or three, if fewer snapshots exist), and then we spline this mass accretion history to obtain a fine-grained mass evolution as a function of redshift. Moreover, in order to avoid spurious identifications, we require that each object has existed as either a central or a satellite for at least three consecutive snapshots. The halo formation time $t^{\rm x}_{\rm form}$ is the earliest moment in cosmic time at which the splined total mass accretion history reaches ${\rm x}\%$ of the peak mass of a halo (we usually express it here in terms of lookback time from the present day, in Gyr). The halo mass is the maximum mass value reached along the main branch: for central haloes the peak mass is usually very close to their mass at $z=0$ and provides a reasonable approximation of the virial mass, generally overestimating it by roughly $10\%$; for satellite subhaloes, the current-epoch mass is usually much lower than the peak mass, because of mass loss due to stripping after accretion onto the parent haloes. This procedure ensures that we have a standard definition of mass and halo formation time that is identical for both central and satellite haloes; however, by construction, subhaloes' ages will always be biased high compared to central haloes' ages.

In what follows, we will adopt various choices for the halo formation time, with, e.g., $t^{\rm 25}_{\rm form}$,  $t^{\rm 50}_{\rm form}$, and  $t^{\rm 75}_{\rm form}$ being the age at which a halo has assembled $25, 50$, and $75\%$ of its peak halo mass. More massive haloes formed more recently than lower mass haloes (see Fig.~\ref{fig:isolation_downsizing}, right panel, black curve; and also e.g. \citealt{Wechsler02}). However, the stellar populations of galaxies residing in more massive haloes at $z=0$ are \textit{older} than the stellar population of galaxies residing in less massive haloes (Fig.~\ref{fig:isolation_downsizing}, right panel, cyan curve), or equivalently, redder (orange curve -- see e.g. \citealt{Heavens04, Thomas05, Nelan05, Jimenez05} for the first observational claims of archaeological downsizing).

\section{Galactic and Halo Conformity}
\label{sec:conformity}
\subsection{Galactic Conformity}

\begin{figure*}
	\includegraphics[width=1.0\linewidth]{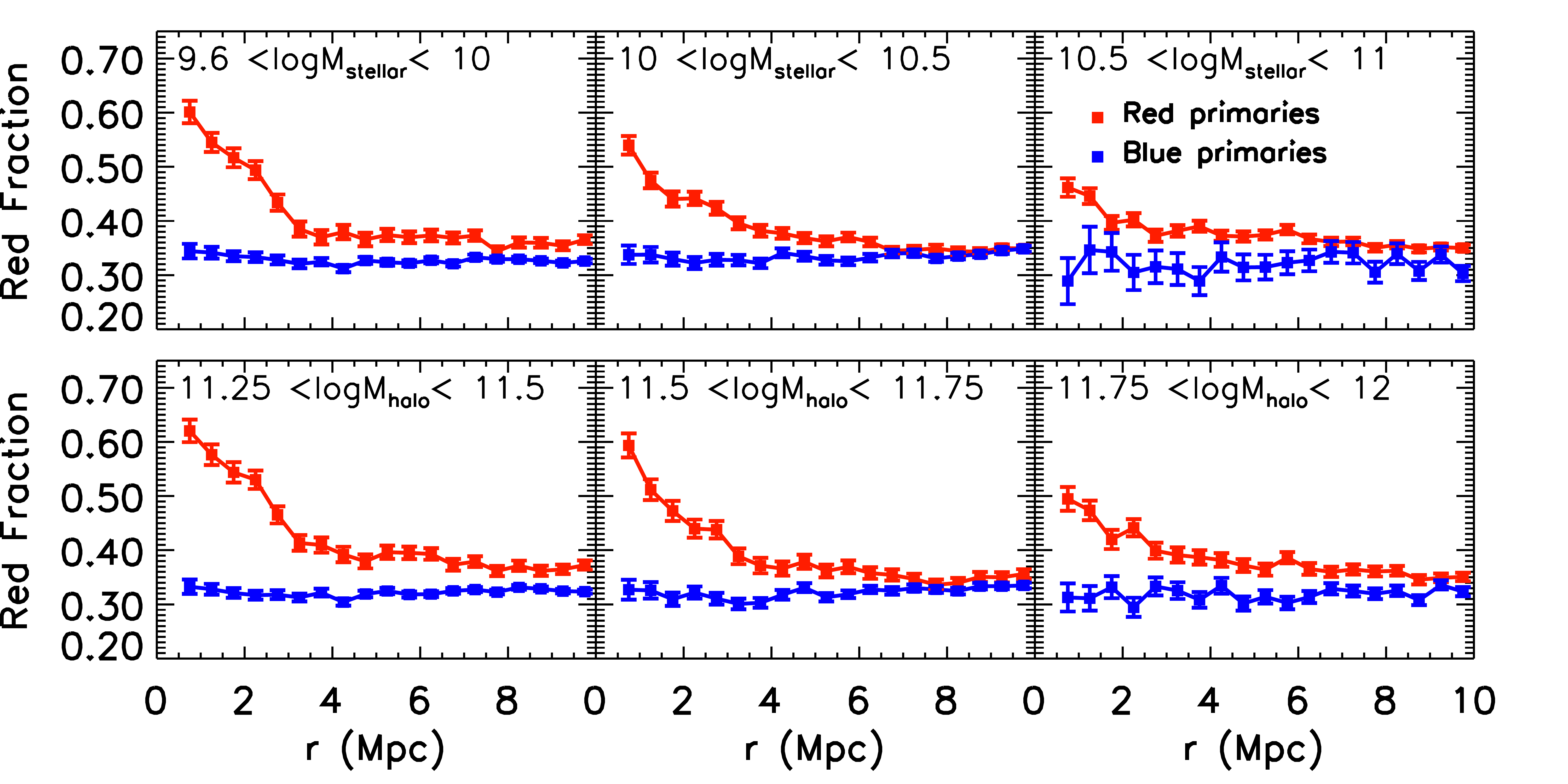}
 	\caption{
        Galactic Conformity in Illustris, i.e. mean red fraction of neighboring galaxies as a function of real-space distance in 500 kpc bins around primary galaxies divided in three different stellar (halo) mass bins in the top (bottom) rows. The red solid line is the fraction of red galaxies around the red population of primary galaxies, while the blue solid line is the fraction of red galaxies around the blue population of primaries. The galactic conformity signal is present at all distances, but it is particularly strong at $r < 3$ Mpc. Standard errors on the means are calculated from bootstrapping. Here the separation between red/quenched and blue/star-forming populations is based on a color cut (see Section \ref{sec:defs}). Moreover, the primaries are isolated galaxies which are mostly centrals, with roughly 20\% fraction of satellite intelopers (see Fig. \ref{fig:isolation_downsizing}); secondaries can be either centrals or satellites, isolated or not.
                 }
 	\label{fig:smconformity}
\end{figure*}

We now present the measurements of galactic conformity in the Illustris simulation. In Fig.~\ref{fig:smconformity}, we plot the mean red fraction of secondary galaxies around their isolated primaries (see definitions and methods in Section \ref{sec:defs}). Each bin is a spherical shell of width 500 kpc, centered on real-space distances from $r=0.75$ Mpc to $r=9.75$ Mpc. Red and blue squares are used to represent red and blue primaries, respectively, as defined in Section~\ref{sec:methods}, and the error bars define the standard error on the mean, as determined from 1000 bootstrap resamplings. The top panel shows the results in three bins in stellar mass for the primary sample, while the lower panel shows the results in three bins in total halo mass. The red (blue) primaries, by stellar mass bin, low to high, have 655 (2921), 1114 (1639), and 1171 (191) galaxies, respectively. The red (blue) primaries, by halo mass bin, low to high, have 559 (3195), 602 (1690), and 721 (654) galaxies. Note that secondary galaxies can have any mass in all panels.

We clearly see that red primaries have a higher fraction of red neighbors than their bluer counterparts. Furthermore, we see two significant trends. First, there is a near-field, higher amplitude conformity signal out to roughly 3 Mpc, and then a plateau of a far-field effect that extends out to at least 5 Mpc. Second, lower mass primaries have both a higher amplitude conformity effect, in both the near- and far-fields, and the far-field effect continues out to larger radii, remaining present out to 10 Mpc in the lowest mass bin. We have confirmed that the signal disappears entirely by 15 Mpc in all cases. We note that for such low mass galaxies, the virial radii of the primaries are significantly lower than even the radii at which we see the near-field effect (the typical viral radii spanning from 150 to about 370 kpc across the three adopted mass bins).

One possibility for the presence of the conformity signal is that red and blue galaxies, selected in fixed stellar mass bins, are nonetheless hosted by halo masses of substantially different size. In this case, the higher red fraction around red primaries would be due to a halo-mass quenching effect. However, as shown by the lower panel of Fig.~\ref{fig:smconformity}, selecting our primary galaxies in halo mass bins does not reduce the amplitude of our signal. Thus, clustering observations that use stellar mass as a proxy for halo mass are unlikely to be biased by this selection technique. We return to the role of mass in the conformity signal in Section \ref{sec:discussion}.

We have also tested these results for robustness to different definitions of quenched/red galaxies. We have used both sSFR and stellar ages, in which quenched galaxies are defined either as being below the median sSFR or above the median stellar age; as having sSFR lower than finite values across stellar mass (e.g., $3, 6, 8 \times 10^{-11} \, {\rm yr}^{-1}$); or in which quenched galaxies are defined using sSFR or stellar age, but at a cut-off to reproduce the same red to blue ratio we have in our fiducial definition. In all cases, a galactic conformity remains, including the near- and far-field distinction: namely, the specific cut to separate red vs blue galaxies does not affect whether there exists a statistically significant signal in a particular mass bin. However, the exact location of, for example, a flat cut in sSFR may affect the presence and magnitude of a dependence of the conformity signal on primary mass. For example, choosing a constant cut in sSFR that is unreasonably high would cause some less massive blue galaxies that are part of ``blue cloud" in Illustris to be classified as red, which would inflate the overall red fraction around more massive galaxies. In relation to this consideration, we argue that a reasonable color cut should follow the general number density contours in color-luminosity space, as opposed to, for example, splitting the blue cloud. Yet we caution that whether or not the conformity signal exhibits a trend with primary mass may depend on the adopted split between star-forming and quenched galaxies.

\begin{figure*}
	\includegraphics[width=1.0\linewidth]{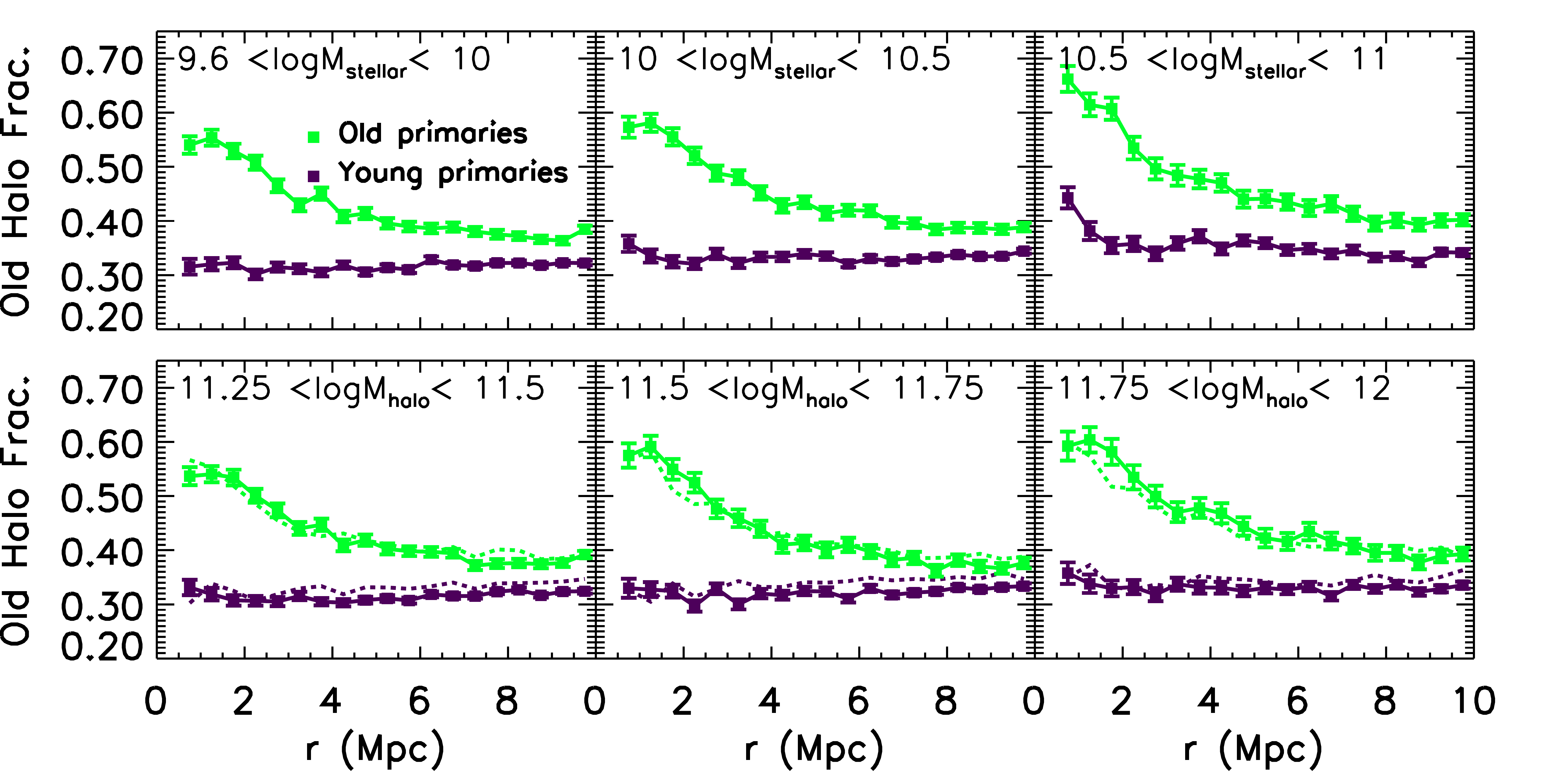}
 	\caption{
        Halo Conformity in Illustris. We show how the effect in Fig.~\ref{fig:smconformity} is paralleled by the conformity between old dark matter haloes. We keep the same isolation criterion, but now, instead of looking at the red fraction around red (blue) galaxies, we plot the fraction of old haloes around old (young) haloes (where halo age is based on $t^{\rm 75}_{\rm form}$). We see the same qualitative signal as before. The split between old and young haloes is defined such that the number of old haloes is the same as the number of red galaxies; this ensures that the relative amplitudes of the conformity signals are the same. In the top row, the stellar mass cuts are for the galaxies that are in each subhalo. In the bottom row, we also overplot as dotted lines the results from Illustris-Dark-1, the dark matter only run with identical initial conditions (without error bars, for clarity). The minimum subhalo mass of secondary subhaloes used to measure the conformity signal in the Illustris-Dark-1 run is $8.8 \times 10^{10} \msun$, and we set the division between old and young subhaloes to match the ratio in the Illustris-1 hydrodynamical run.
                }
 	\label{fig:dmonly_conformity}
\end{figure*}

\subsection{Dark Matter Halo Conformity}

In Fig.~\ref{fig:dmonly_conformity}, we now present the dark-matter halo conformity effect in the Illustris Simulation. In order to be able to compare the relative amplitude of the halo and the galaxy conformity effects, we split Illustris haloes into two samples of haloes, old and young, which are chosen to match the sample sizes of the red and blue populations. Here we adopt the $t^{75}_{\rm form}$ definition for halo formation time. In the case of Illustris-1, the global split between old and young haloes occurs at $z = 0.965$.

We keep the same isolation criterion as before defined by stellar mass, but now the binning of primary galaxies is by halo mass. In this way, we check only for the influence on the conformity signal of using stellar versus halo mass, rather than introducing another difference due to a new isolation criteria. For the top panel, the stellar masses refer to the galaxies that inhabit those haloes, as we make the measurements in the fiducial Illustris-1, so haloes and galaxies have direct counterparts.

A dark-matter halo conformity signal is detected. Old haloes are preferentially surrounded by other old haloes. Similarly, there is the same near- and far-field split as in the galactic effect, namely the halo conformity signal is strongest at separations $ < 3$ Mpc. Halo assembly bias, and its effect on clustering, has been studied for many years in other dark matter-only simulations and with semi-analytic models (e.g., \citealt{Gao05}; \citealt{Wechsler06}; \citealt{Croton07}; \citealt{Dalal08}; \citealt{LP11}; \citealt{Wang13}). Thus, it is not particularly novel that we find such a signal in Illustris. What is notable is that, as the Illustris Simulation has full baryonic physics, we can compare the qualitative shape of the galactic conformity, as seen in Fig.~\ref{fig:smconformity}, with the halo conformity as seen here in Fig.~\ref{fig:dmonly_conformity}.

Overall, the galaxy and halo conformity signals are qualitatively similar. Both show a higher amplitude effect out to $\sim 3$ Mpc, and the signal plateaus at larger radii. One difference is that the halo conformity signal shows no obvious decline with increasing halo mass. At first glance, this is in contradiction with the mass dependence seen first in \citet{Gao05}, in which the relative bias of the oldest to the youngest haloes was a function of mass. That is, the relative bias was found to be higher at lower masses. Three differences likely account for this apparent discrepancy. First, the range of halo mass that we probe directly in this analysis, from $\rm{log}\,\mhalo = 11.25$ to $\rm{log}\,\mhalo = 12$, is much smaller than that probed by \citet{Gao05}. Our mass range corresponds roughly to the four lowest mass bins in Fig.~3 of that work, over which the relative bias is nearly constant. Any remaining difference might be attributable to the facts that a) our split between young and old haloes is global and not halo mass dependent; and b) the bias there is measured between $6 \hmpc$ and \,$25 \hmpc$, whereas most of our measurements lie at smaller radii, $\lesssim 10$ Mpc.

Another difference between the galactic and halo conformity signals is that the decline in old fraction with radius is more gradual than the decline in red fraction, with the sharp feature seen especially prominently in the lowest mass bin not present in Fig.~\ref{fig:dmonly_conformity}. This suggests that there could be a particular scale at which quenching becomes efficient. This could be related to internal processes, such as stellar or AGN feedback, akin the one halo effect offered by \citet{Hartley15}, or it could be due to external processes, such as the influence of the tidal radii of neighboring haloes \citep{Dalal08, Hearin15}. Finally, for the halo conformity signal, the old fraction around young and old haloes does not return to the global average until almost 15 Mpc, whereas in the galactic conformity case, the fractions are equal by 6 to 10 Mpc, depending on mass bin.

These results are robust to the effect of baryonic physics: indeed, we confirm the presence of an equivalent halo conformity signal in the Illustris-1-Dark simulation. This is overplotted for old and young primaries in the lower panels of Fig.~\ref{fig:dmonly_conformity} as dashed lines (green and purple, respectively). The Illustris-1-Dark simulation was run with the same initial conditions and box size as Illustris-1, but with only dark matter. Because this run does not have stellar masses, we cannot keep the identical isolation criteria for this test. We instead define a halo mass cut of $8.8 \times 10^{10} \msun$, which is more than two times below the minimum primary halo mass in our sample. We can thus apply an equivalent isolation criterion to our dark matter haloes -- we require that primary haloes have no neighbor within 500 kpc that has more than half the primary's halo mass -- and obtain a similar sample size with which to compare results between Illustris-1 and Illustris-1-Dark. We set the division between old and young haloes to match the global ratio of old to young haloes in the Illustris-1 run. 

The results show that the halo conformity signal is also present in the dark matter-only run. Specifically, the shape of the halo conformity signal is essentially identical between the Illustris-1 and Illustris-1-Dark runs. We thus find that this measure of halo assembly bias is not affected by the inclusion of baryons via hydrodynamics and feedback.


\section{The Color--Age Relation}
\label{sec:colorage}

Having seen that Illustris exhibits both a galactic and a halo conformity, and that they appear to be similar in qualitative shape and amplitude, we turn our attention to the connection between the two. In this Section, we emphasize the relationship between galaxy color and halo age, which may explain how the baryonic signal arises from the dark matter one. As \citet{Sparre15} showed for central galaxies in Illustris, at least some measures of formation time (in that case, when half the stellar mass is formed) correlate well with galaxy color. In this work, we examine formation times based on the build-up of the full halo, including mass from dark matter, stars, and gas. The total mass is chosen because the underlying ansatz of basic HOD models is that mass is the only contributing factor to galaxy occupation. By adding in the halo formation time, we simply add one additional parameter, which attempts to encompass the time evolution of that halo mass. We do not argue that any particular formation time is \textit{a priori} superior, but we will show subsequently that not only are the correlations different, but some do a better job at reproducing the measured two-point statistics.

\begin{figure*}
	\includegraphics[width=1.0\linewidth]{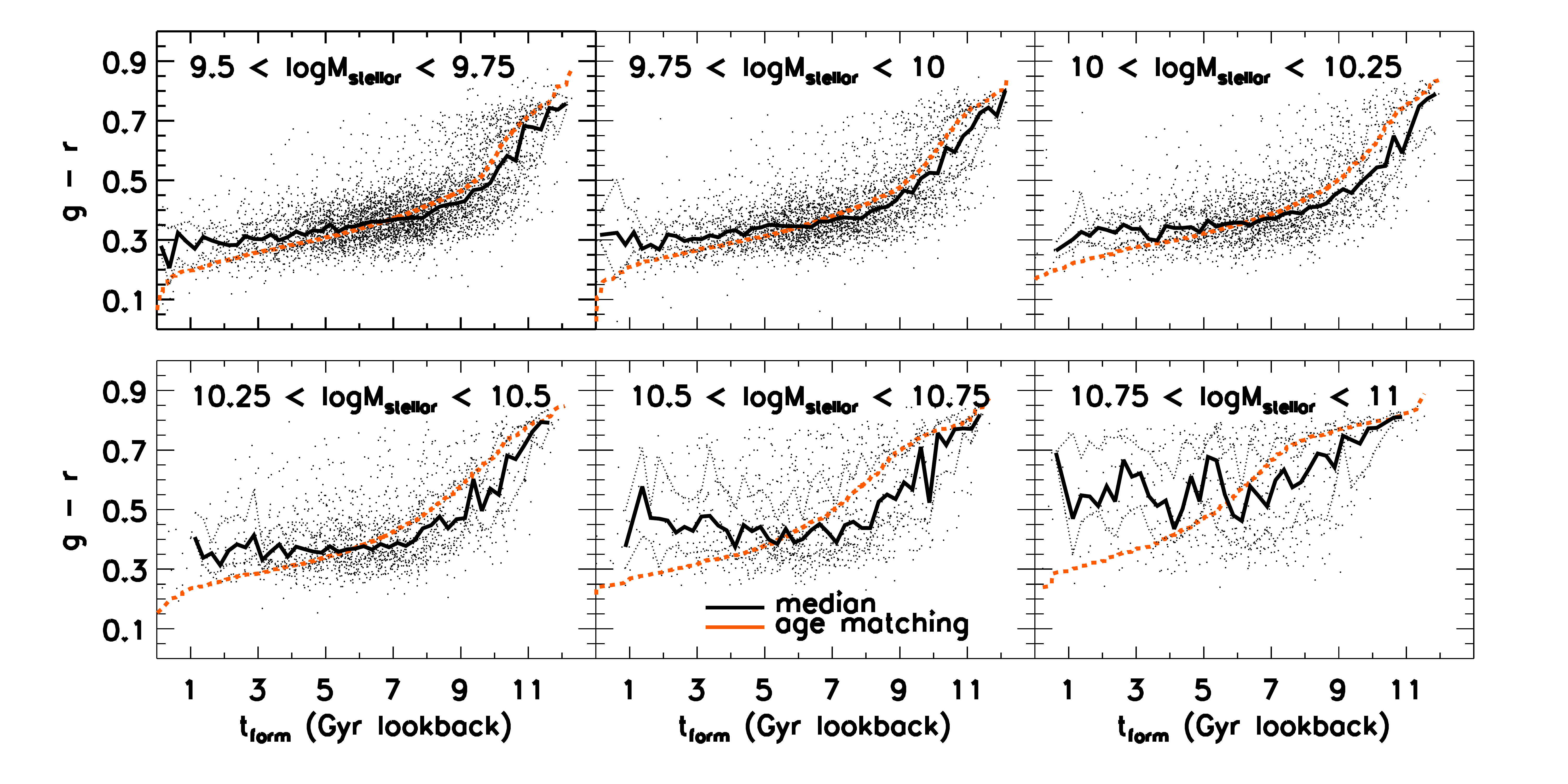}
 	\caption{
	Color--Halo Age Relation in Illustris. We show the galaxy color versus halo formation time in bins of stellar mass. Here the halo formation time is defined as the accretion time of $75\%$ of the maximum mass for both centrals and satellites (see Section~\ref{sec:mergertree} for details.) Black points are each galaxy in a given stellar mass bin. The solid black line is a running median of the black points. The dashed orange line defines the monotonic relationship between color and halo formation time, if we rank order the simulated galaxies by assigning the reddest galaxies to the oldest haloes. We use this model in Section~\ref{sec:discussion} to reproduce the conformity results from the hydrodynamical simulation. 
        }
 	\label{fig:colorage}
\end{figure*}

In Fig.~\ref{fig:colorage} we show that this correlation between galaxy color ($g - r$) and halo age, $t^{75}_{\rm form}$ exists in the Illustris Simulation. We see that older haloes tend to host redder galaxies. However, the trend is split into two distinct regions. Namely, the reddest galaxies exist almost exclusively in the oldest haloes, but the reverse is not true; there are old haloes that contain bluer galaxies. This creates a sharp break in the correlation. As can be seen by the running median (solid black line), in the lower stellar mass bins, however, there is still a weak correlation between halo age and color even among the bluer galaxies.
The trend is no longer visible at increasingly high stellar masses. In the two highest mass bins, there is substantial scatter, with a population of red galaxies hosted by young haloes. The halo masses of galaxies in these stellar masses range from roughly $3.3 \times 10^{11} {\rm M}_{\odot}$ to $1.25 \times 10^{13} {\rm M}_{\odot}$. Especially at the high halo mass end, internal quenching may be responsible for destroying the tighter correlation seen at lower stellar masses. In particular, radio-mode feedback from the AGN will be a contributing factor to the color of the galaxies.

To guide interpretation, we also show the maximal correlation between galaxy color and halo age (orange dashed line). This is similar to the ``age matching" of \citet{Hearin14}. For each 0.25 dex stellar mass bin, we rank order the galaxies by our halo age. We then assign the reddest galaxy in that stellar mass bin to the oldest halo in that stellar mass bin. We continue this process, such that the bluest galaxy is eventually assigned to the youngest halo in this mass bin. We then repeat this process for each mass bin. If the relationship between color and halo age were in fact monotonic, with redder galaxies always present in older haloes, we would expect the galaxies in Illustris to lie along the overplotted, orange dashed line. Instead we see that, due to scatter in the correlation, older haloes are bluer and younger haloes are redder than the most simplistic model would predict.

Satellite galaxies serve to enhance the overall trend in the color-halo age plane. By construction, and because of their infall times, they tend to populate the older-halo end rather than the young one. Because they also have a higher red fraction at fixed halo (or stellar) mass, this drives the sharp upward shape of the correlation. Certainly, though, the functional form and scatter at fixed halo age of the color--age relation depends on the exact choice of the definition for halo formation time: with $t^{75}_{\rm form}$, for example, values necessarily extend to more recent halo formation times than the analog $t^{25}_{\rm form}$ values. An equivalent relation exists in Illustris between halo age and sSFR, and halo age and stellar age. 

\section{Discussion}
\label{sec:discussion}

\subsection{Modeling and Interpretation}
\label{subsec:modeling}

The parsimonious explanation for why the halo and galaxy conformity signals are qualitatively the same is to invoke a correlation between halo age and galaxy color. If such a relationship holds, then in the presence of dark-matter halo conformity based on halo formation time (i.e., halo age), we would expect a color-dependent galaxy conformity signal to necessarily arise. In the previous sections we have shown that this is indeed the case in the Illustris hydrodynamical simulation, with the galactic conformity signal resulting from the combination of the aforementioned two effects.

First, initial conditions, plus time-evolution under gravity, produces halo conformity. Studies in dark matter only simulations (\citealt{Keselman07}; \citealt{Dalal08}; \citealt{Hearin15}) provide possible physical mechanisms for this effect that we do not explore, since we are concerned primarily with the role played by baryons. Second, due to the correlation between stellar mass and dark matter mass build-up, Illustris naturally exhibits a color--halo age relation at $z=0$. Together, they produce a galactic conformity effect. 

\begin{figure*}
	\includegraphics[width=1.0\linewidth]{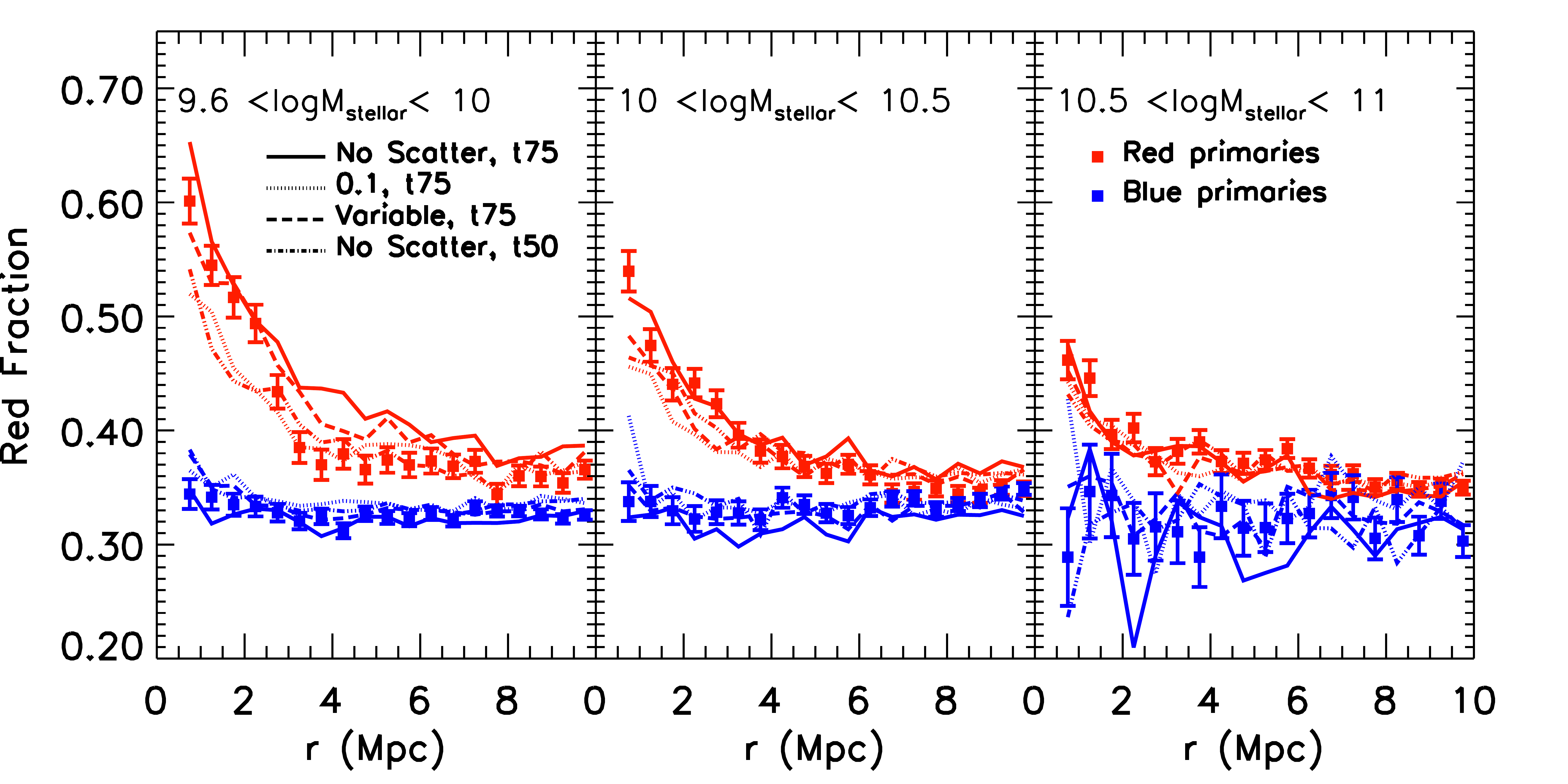}
 	\caption{
        Galactic Conformity vs Semi-Empirical Modeling. Data points with errors are Illustris measurements from Fig.~\ref{fig:smconformity}, while dashed curves represent the results from our semi-empirical abundance + age matching model, using two different definitions of halo age: $t^{75}_{\rm form}$ and $t^{50}_{\rm form}$. We also show the effect of adding scatter to the color--halo age relation for use in the age matching. More recent definitions of halo age produce larger conformity amplitudes at all separations, while adding scatter to the color--halo age relation lowers the amplitude of the conformity signal.}
 	\label{fig:modelconformity}
\end{figure*}

To test whether this qualitative picture works in practice, we present in Fig.~\ref{fig:modelconformity} the measured galactic conformity signal with the same method as used in Fig.~\ref{fig:smconformity}, except that now we have reassigned the galaxies to different haloes according to a toy semi-empirical model. 
The fiducial model (`No Scatter, t75') works as follows. First, we assign galaxies to dark matter haloes via subhalo abundance matching (\citealt{Conroy06}; \citealt{VO06}), with a scatter of $\sigma_{M} = 0.18$, which falls in between various literature values (e.g., \citealt{Han15}; \citealt{ZM15}) and which is roughly consistent with the actual scatter at fixed halo mass in the stellar-to-halo mass Illustris relation (this varies within 0.13-0.23 over the range $\rm{log}\,\mhalo = [11.3,13.5]$). Next -- following the same basic procedure as \citet{HW13}, but using our $75\%$ formation time, $t^{75}_{\rm form}$ -- in bins of 0.25 dex in \rm{log} $\mstars$, we rank galaxies according to color, and we assign the reddest galaxy to the oldest halo. This would be equivalent to the orange line in Fig.~\ref{fig:colorage}, except that, rather than using stellar masses directly from Illustris, we first perform subhalo abundance matching. 

The results from the model are shown as solid lines in Fig.~\ref{fig:modelconformity}, overplotted to the points from Fig.~\ref{fig:smconformity}. It bears mentioning first that when we perform only abundance matching -- and no age matching of any kind -- and thus erase any connection between color and halo age, we obtain a null result, in which the red fraction is indistinguishable for red and blue primaries in all mass bins. On the other hand, when we add in age matching, we see that the results are qualitatively the same as those found directly from the hydrodynamical simulation, with some small differences especially at the lowest mass bin. Indeed, we know from Fig.~\ref{fig:colorage} that the color--age relation in Illustris is not, in fact, strictly monotonic and exhibits a large scatter in color at fixed halo age: therefore, our fiducial model is implementing an extreme version of a color--halo age relation. 
 
As we noted in Section \ref{sec:colorage}, the functional shape of the color--age relation depends on the exact choice for the definition of halo age. We have therefore checked the output of the above-described semi-empirical model for different definitions of halo formation time. Moreover, we have tested the dependence of the modeled conformity signal also on the adopted amount of scatter in the color--age relation. 

Three of these additional models for different choices of formation time and scatter in the color--age relation are shown for comparison and insight in Fig.~\ref{fig:modelconformity}. In all three, the subhalo abundance matching remains fixed. The `0.1, t75' model adds 0.1 Gaussian scatter in $g - r$ at fixed halo age. The `No Scatter, t50' model has no scatter, but ranks galaxies according to halo age as defined by the $50\%$ formation time. Finally, the `Variable' model has weaker scatter at low stellar mass, and increasingly strong scatter at high stellar mass. Specifically, using the same 0.25 dex bins that we use to assign colors, we increase the scatter by 0.02 per bin, beginning with 0.02 scatter in the $\rm{log}\,\mstars = [9.25, 9.5]$ bin. Thus, this model has 0.1 scatter (equivalent to the `0.1, t75' model) in the $\rm{log}\,\mstars = [10.25, 10.5]$ bin, and higher scatter at higher masses than that. These adopted scatter choices encompass the actual values from Fig.~\ref{fig:colorage}, even though with somewhat different functional dependences on halo age and mass.

The models all have quite similar consequences for the conformity signal in the two larger primary stellar mass bins. However, for the lowest stellar mass bin, the `No Scatter, t50' and `0.1, t75' models both under-predict the conformity signal below 3 Mpc. Meanwhile, both the `No Scatter, t75' and the `Variable Scatter, t75' models over-predict the conformity signal beyond 3 Mpc. 

More importantly, two general trends can be taken away from such a comparison. First, more recent definitions of halo age produce larger conformity amplitudes at all separations. Compared to $t_{\rm {form}}^{\rm{75}}$, using $t_{\rm {form}}^{\rm{25}}$ (not shown) lowers the amplitude of the signal significantly, while using $t_{\rm {form}}^{\rm{50}}$ lowers it by an intermediate amount. No particular formation time measure is \textit{a priori} better than any other, but in the Illustris Simulation, $t_{\rm {form}}^{\rm{75}}$ leads to the greatest spread in halo ages for central galaxies. We argue that this allows for a less noisy relationship between halo ages and galaxy colors, and therefore results in a larger inferred galactic conformity signal (at least for the specific color-based cut adopted in this work).
Second, as should be expected, adding scatter to the color--halo age relation lowers the amplitude of the conformity signal. This trend continues with higher values of the scatter until the entire signal is washed away. The implication of this is substantial: galactic conformity cannot exist without a reasonably tight relationship between the galaxy property (in this case, color) and the underlying dark matter property (in this case, halo age), which itself is clustered. The fundamental clustering is between dark matter haloes; galactic conformity comes along for the ride.

\begin{figure*}
	\includegraphics[width=1.0\linewidth]{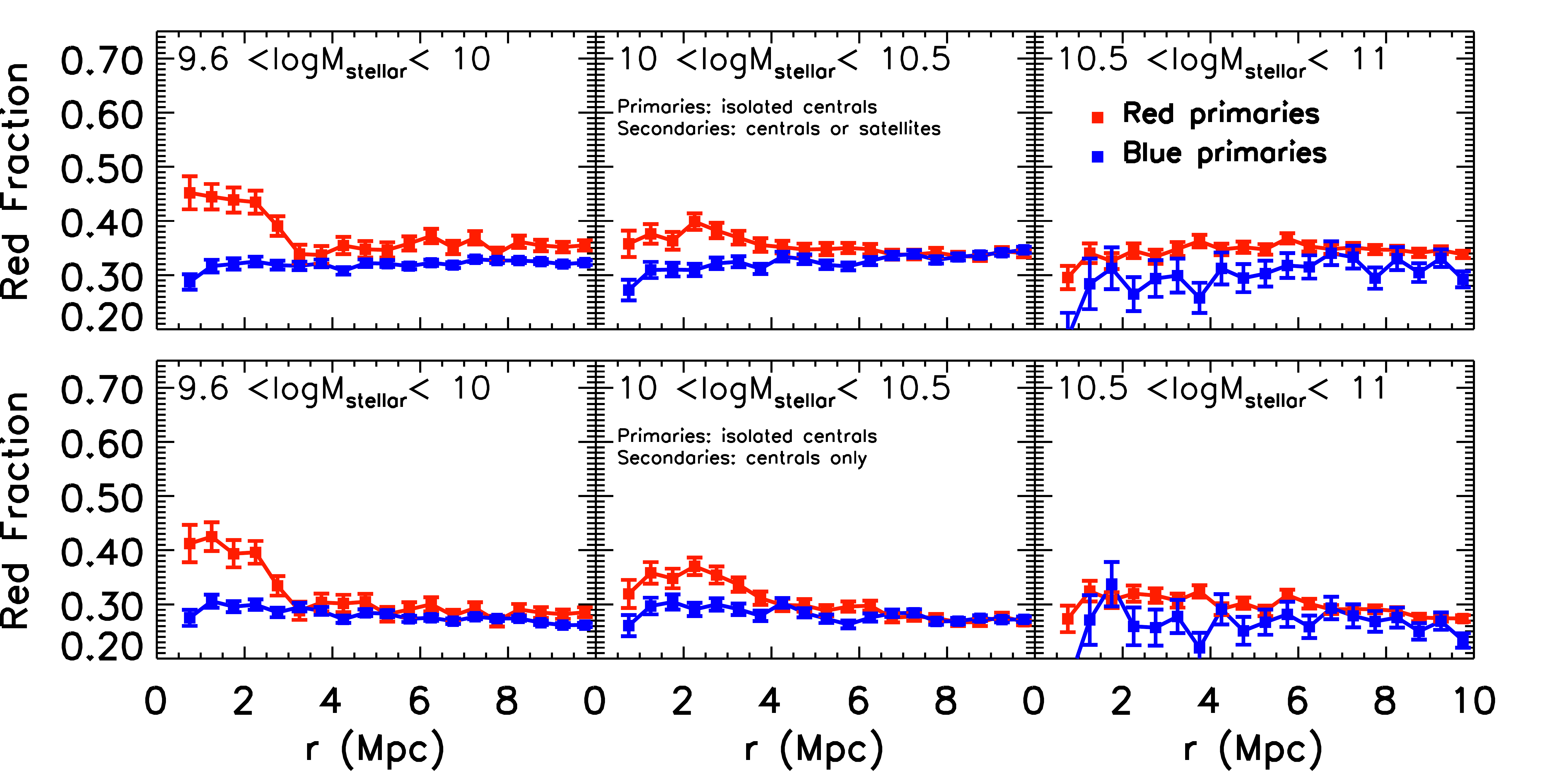}
 	\caption{
	Galactic Conformity in Illustris for different galaxy selections.
       Top Panel: We show the conformity signal as calculated selecting only central galaxies, in addition to the isolation criterion, in stellar mass bins. A conformity signal remains at $r < 3 \, \textrm{Mpc}$, but the signal out to 10 Mpc is substantially reduced. Thus, while the isolation criterion selects mostly central galaxies for the primaries, much of the observed conformity signal at larger scales appears due to the correlation between low-mass primaries and satellites around other central galaxies. Bottom Panel: We show the conformity signal as calculated using only central galaxies, for both the primary sample and the secondary neighbors. With no satellites included in either sample, a conformity signal remains at $r < 3 \, \textrm{Mpc}$, but the signal out to 10 Mpc is completely eliminated.
       }
 	\label{fig:centrals_only_primary}
\end{figure*}

\subsection{The Roles of Centrals and Satellites}
\label{subsec:satellites}

In presenting our results, we have measured the galactic conformity signal of all neighboring galaxies above a stellar mass threshold around isolated primary galaxies. However, a subset of satellite galaxies remains in the primary sample despite the isolation criterion. Furthermore, the secondary sample includes \textit{all} galaxies, regardless of whether they are central or satellite. In this Section, we present the effect of satellites in Illustris on the galactic conformity signal, with its possible implications for observational searches for assembly bias.

In Fig.~\ref{fig:centrals_only_primary}, we show two analog versions of the measurements presented in Fig.~\ref{fig:smconformity} but for different selections of the primary/secondary samples. In the top panels, we require the primary galaxies not only to be isolated, but also to be centrals, as defined in the Illustris catalogs. That is, the galaxy must belong to the central {\sc subfind} halo in a given {\sc fof} halo, \textit{and} it must still be isolated, according to our previous definition. We thus ensure that we are not introducing any added effect from nearly equal mass, merging subhaloes, and remove the effect of the 10\% -- 20\% of satellites that would otherwise remain in the primary sample (see Fig.~\ref{fig:isolation_downsizing}).
Compared to Fig.~\ref{fig:smconformity}, we see that the quenched fraction around red galaxies is reduced in the near-field (out to 3 Mpc) by about a third in the lower and middle stellar mass bins, and almost eliminated in the highest mass bins. Moreover, the signal amplitude beyond 3 Mpc is also reduced, although as can be seen for the lowest mass bins, it remains statistically significant out to 10 Mpc nonetheless. 

The influence of satellites becomes enlarged when the isolation (or central) criterion is extended to the the secondary neighbors. If one is interested primarily in the galaxy assembly bias within central haloes, then this is the best way to control for any possible satellite effect. In Fig.~\ref{fig:centrals_only_primary} (bottom panel), we have thus limited our measurement to only isolated, central galaxies for the \textit{primaries} and central galaxies for the  \textit{neighboring secondaries.} While a conformity signal now remains out to $\sim 3 $ Mpc -- one which is nearly identical to the top panel -- the far-field conformity effect between 3 -- 10 Mpc has completely been suppressed. In addition, the overall quenched fraction has also been reduced, as satellites are preferentially redder than central galaxies.

In Illustris, we thus find that while satellites have a significant role in the emergence of a conformity signal, central galaxies \textit{on their own} can produce a signal beyond the virial radius and out to roughly 3 Mpc in real-space distance. Therefore, it is possible that repeated analyses of observations using different methods for classifying isolated galaxies will obtain somewhat different quantitative results. 

Interestingly, these results from Illustris imply that the terms 1-halo and 2-halo used to describe the two conformity regimes may be misleading. Specifically, in Illustris, we find a conformity signal out to 3 Mpc that, while enhanced by the inclusion of satellites, is still present when using only centrals, at least for lower primary masses. Thus, the near-field effect is in fact a 2-halo correlation. Conversely, eliminating satellites entirely from both the primary and secondary samples, as opposed to just the primary sample, eliminates the far-field signal beyond 3 Mpc. Thus, this portion of the signal in Illustris appears due to the correlation between centrals and satellites around \textit{other} centrals, a 2-halo term, but one that will be diminished if there is no 1-halo correlation between centrals and their own satellites.

We postpone to future analysis the task to identify the physical mechanisms responsible for the emergence of the near- and far-field effects. In fact, whether or not it makes sense to distinguish between centrals and satellites in understanding assembly bias depends of the question being asked, but as noted by  \citet{Hearin15}, many satellites may have only recently been centrals themselves, and likewise, centrals may have been formerly satellites during close interactions.


\subsection{Tests of Mass-Dependence and the Central-Satellite Split}

\begin{figure*}
	\includegraphics[width=.45\linewidth]{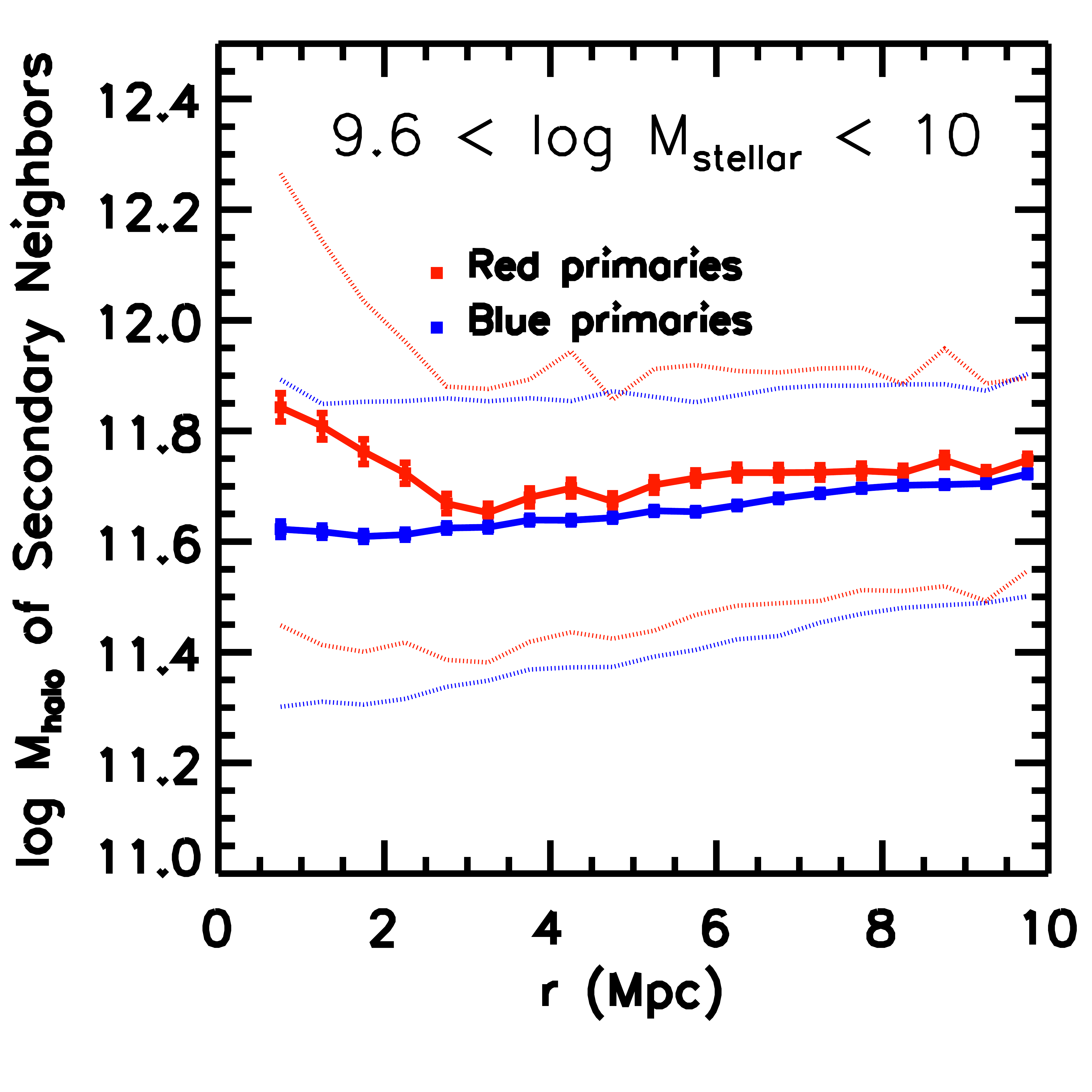}
	\includegraphics[width=.45\linewidth]{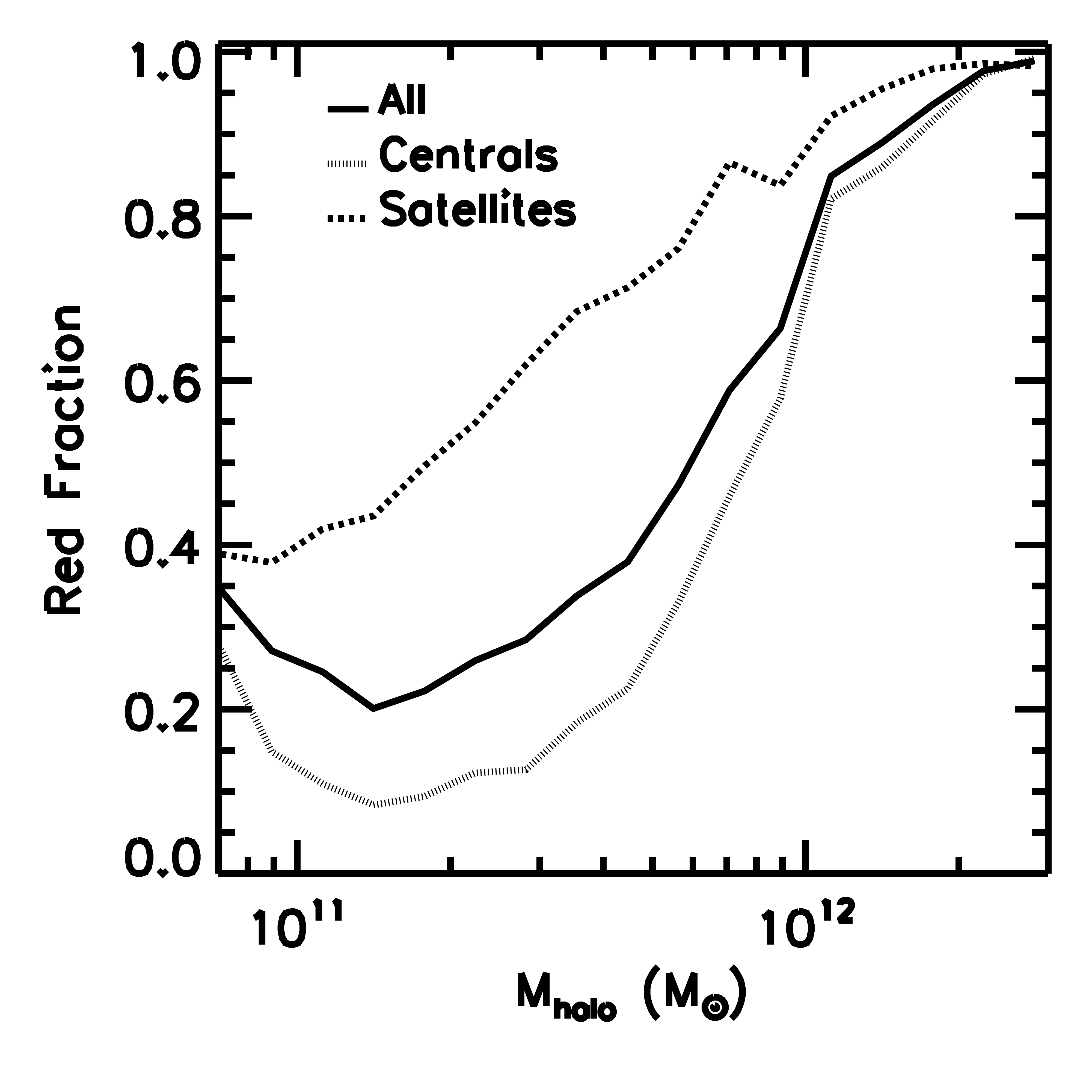}
	\includegraphics[width=.45\linewidth]{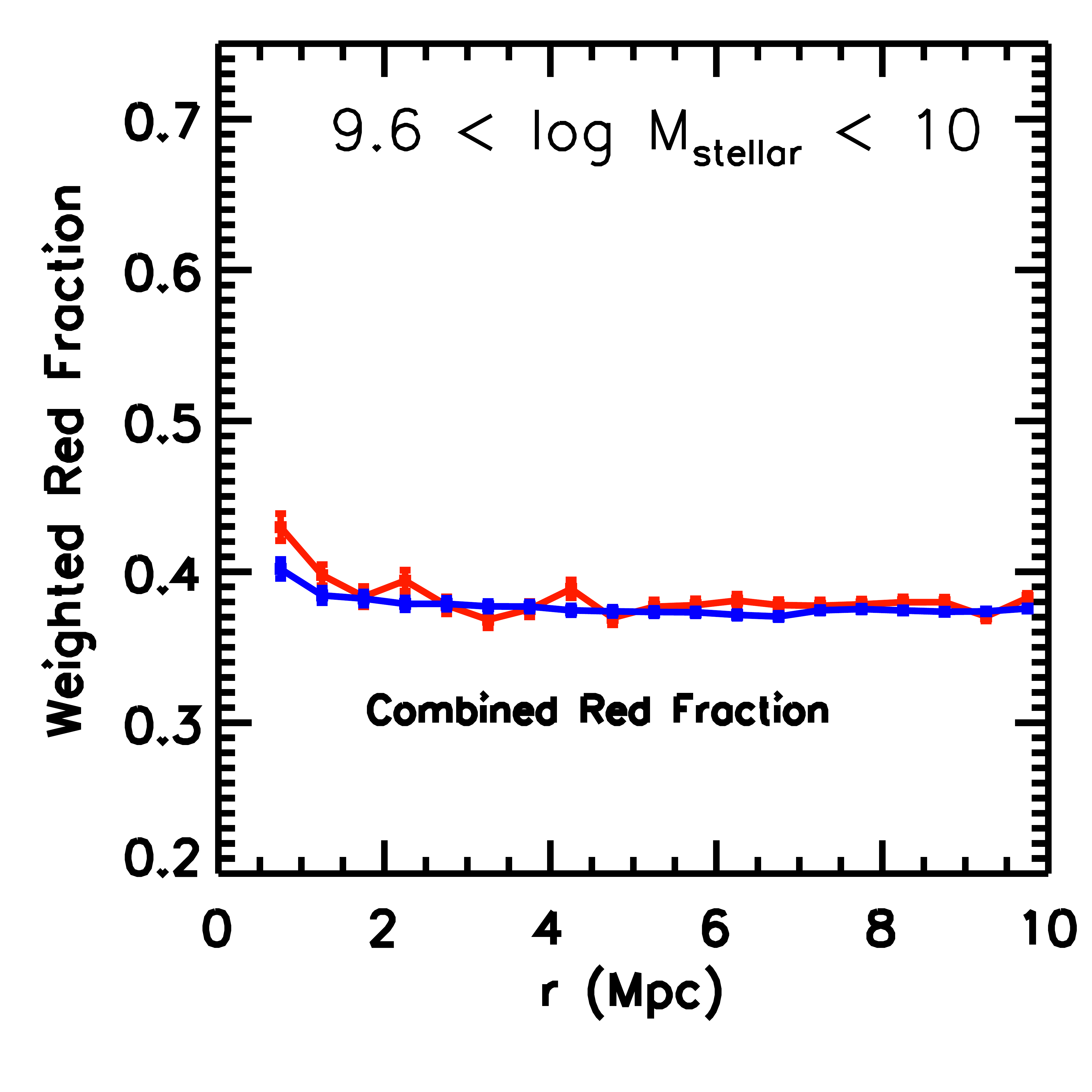}
	\includegraphics[width=.45\linewidth]{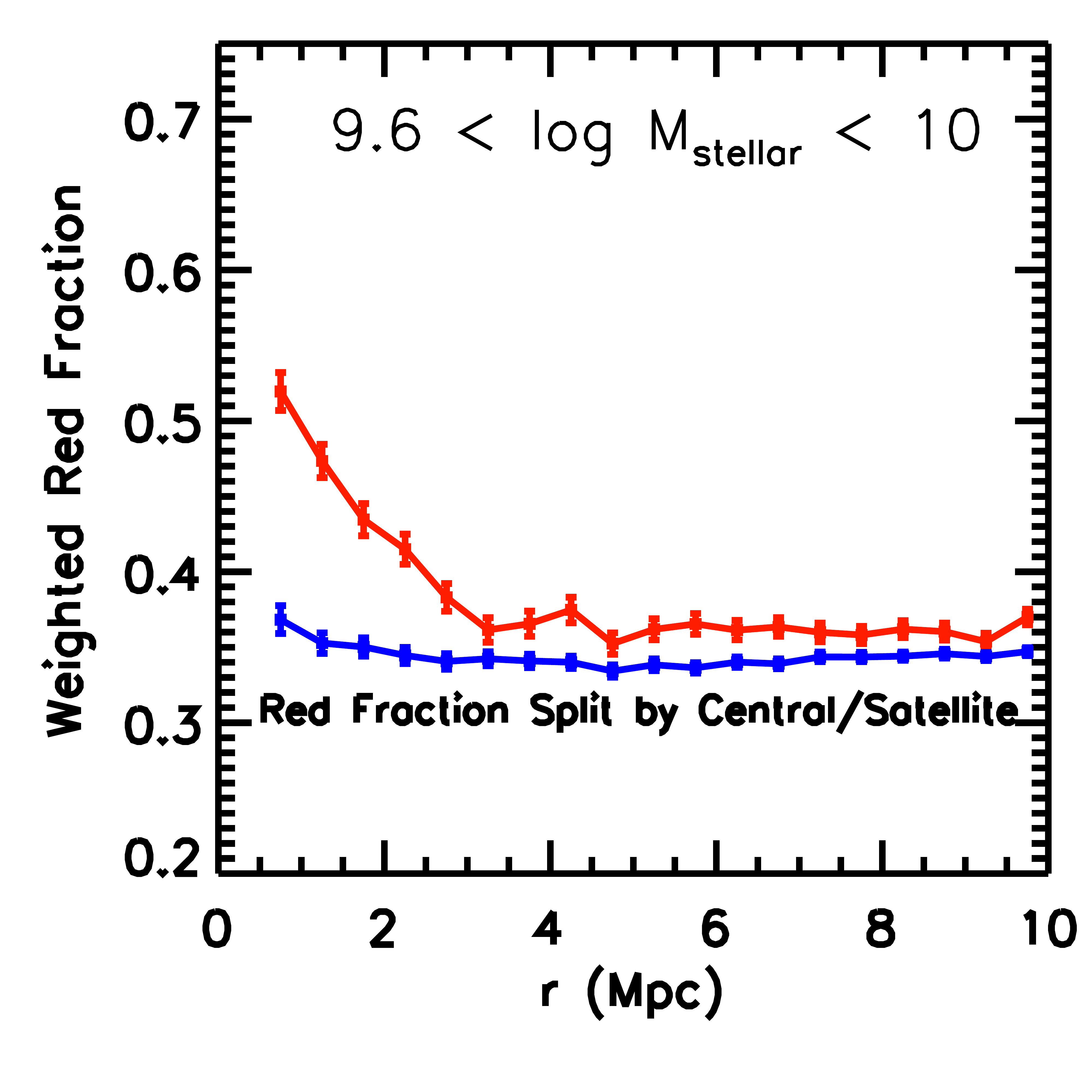}
 	\caption{
	In the top left panel, we show the mean halo mass, along with the 25th and 75th percentiles, of the neighbors around primary galaxies in the $9.6 < \rm{log} \, \mstars < 10$ stellar mass bin. We choose this bin for demonstration purposes, because the difference between red and blue galaxies is maximal compared to the other stellar mass bins. (In the largest mass bin, the reverse is actually true: blue primaries are surrounded by slightly more massive galaxies than red primaries.) Because red fraction generally increases with stellar mass (see Fig.~\ref{fig:isolation_downsizing}), we test whether this mass difference is sufficient to account for the conformity signal.
	In the top right panel, we show the red fraction as a function of halo mass for both centrals and satellites, as well as both the joint population. We note that halo mass is defined as the peak mass of the accretion history. We use this relationship to test the effect the halo mass has, on its own, on the conformity signal compared to the effect of using additional properties that are sensitive to assembly history.
	In the bottom left plot, we assign to each galaxy the value of the mean quenched fraction at its halo mass, as defined by a spline over narrow bins in halo mass. We plot the mean of these weights for neighboring galaxies as a function of real-space distance in 500 kpc bins for the $9.6 < \rm{log} \, \mstars < 10$ stellar mass bin of the primary galaxies. If the conformity signal is due entirely to quenched galaxies being surrounded by more massive galaxies than unquenched galaxies, then the signal should be of the same magnitude as in Fig.~\ref{fig:smconformity}, whereas if the signal is not due to the masses of the secondary galaxies, but rather dependent on assembly history, then there should be no signal, because all that information has been discarded by considering only mass. As can be clearly seen, there is no remaining signal.
	In the bottom right panel, we can add back in the dependence on halo assembly, but without directly using $t_{\rm form}$ as a proxy. Rather we weight the centrals and satellites separately, in each case by mean red fraction as a function of peak halo mass, and we qualitatively recover the conformity signal This reinforces the idea that there is nothing special about formation time, per se. Rather, any additional measure of halo assembly, including splitting by central and satellite galaxies, introduces a bias not included by a single definition of mass.
        }
 	\label{fig:wconformity}
\end{figure*}

In Fig.~\ref{fig:wconformity}, top left, we show that for the lowest stellar mass bin, there is a difference in the mean (thick lines), as well as the 25th and 75th percentiles (thin lines) in the masses of the neighbors of red and blue galaxies. It is thus natural to ask whether this difference -- which is maximal for this stellar mass bin, and disappears at the highest stellar mass bin -- is sufficient to reproduce the conformity signal. Given the role that satellite galaxies play in producing the full galactic conformity signal, however, it is also sensible to ask whether the conformity signal can be modeled knowing only that a particular halo is a central versus a satellite, in addition to knowing its mass, rather than the particular epoch of formation. In this scenario, the exact definition of halo formation is unimportant; merely we question whether knowing something of the halo's environment is a sufficient proxy for its assembly history. We tackle these issues in this Section, by further noticing that the red fraction for centrals and satellites as a function of halo mass is substantially different, with satellites being on average much redder than centrals (Fig.~\ref{fig:wconformity}, top right): therefore, even if mass alone is insufficient to explain the conformity signal, the central--satellite division could produce an effect in the correct direction.

In Fig.~\ref{fig:wconformity}, bottom left, we confirm that the clustering dependence we measure in Illustris is not merely a product of low-mass quenched primaries being in the large-scale environment of more massive secondaries. To test this, we measure a weighted conformity signal, in which the weights are assigned solely as a function of the peak halo mass. First, we measure the mean quenched fraction as a function of halo mass in appropriately small bins (Fig.~\ref{fig:wconformity}, solid line, top right.) Then, by fitting a spline over the halo masses our our galaxy sample, we have assigned an average ``quenched weight" to each halo based on its total halo mass. We then measure the equivalent mean weighted quenched fraction as a function of radius.

If the conformity signal was caused by higher red fractions in higher mass bins, combined with the fact that smaller mass quenched galaxies were more often found in the vicinity of these haloes, then we would expect to reproduce our conformity signal from Fig.~\ref{fig:smconformity}. In this case, however, we see in Fig.~\ref{fig:wconformity} (bottom left) that erasing the spatial information and measuring the conformity signal due to the average red fraction completely eliminates all trace of  conformity. This is because it is not merely the halo masses making the galaxies around quenched galaxies preferentially quenched, but some other property associated with halo assembly. However, it also suggests that satellite-specific baryonic processes may be partially degenerate with long-term environmental effects for creating the color--halo age relation. Still, we note that based on the previous subsection, satellite quenching mechanisms cannot be solely responsible, because galactic conformity remains, even when only central galaxies are considered, albeit out to smaller scales.

Now, we add the assembly information: we repeat the above procedure for assigning quenched weights independently for centrals and satellites. Each central is assigned a weight based on its red fraction as a function of central peak halo mass, and each satellite is assigned a weight based on its red fraction as a function of satellite peak halo mass. Once again, we measure the average mean weighted quenched fraction around our primary galaxies. The results (shown again for the $9.6 < \rm{log}\,\mstars < 10$ stellar mass bin; Fig.~\ref{fig:wconformity}, bottom right panel) replicate the qualitative results seen in the top left plot of Fig.~\ref{fig:smconformity}: namely, we can model galactic conformity also by using the central vs satellite split. This is not surprising: satellites are, by construction, significantly older at fixed stellar mass than central galaxies. Thus, using the central vs. satellite information rather than the formation times gets at similar information regarding the assembly history. Both formation time and whether a galaxy is a satellite are proxies for the much more complicated information encoded in the full history of a halo's location within its environment. However, the amplitude of this signal is roughly similar to the `No Scatter, t50' model in Fig.~\ref{fig:modelconformity}. Thus, like the 50\% formation, the central-satellite split contains less information about the color of a galaxy than its 75\% formation time.

\subsection{Toward Comparisons with Observations}
\label{subsec:observations}

In this paper, we have focused on the measurement of the conformity signal in the Illustris simulation almost exclusively from a theoretical perspective. However, in the previous sections we have touched upon typical observational choices that can affect the measured amplitude and shape of the galactic conformity effect. Here, we attempt a summary of such approaches and present future analysis directions.

A few choices distinguish this work from observations-based ones (e.g., \citealt{Kauffmann13}) and make direct comparisons difficult to interpret. These include: 
1) different isolation and selection criteria, which affect satellite galaxy inclusion; 2) 3D {\it vs} 2D projected separations with additional redshift-space cuts; 3) different averaging approaches for the measurement of the signal, i.e. non-stacking {\it vs} stacking techniques of the primary galaxies; 4) different splittings of the galaxy population into quenched and star-forming samples, as fixed binary groupings (red and blue galaxies) {\it vs} a more continuous split by sSFR percentiles typically done in observations; and finally  5) different apertures for the measurement of the relevant galaxy properties themselves, either colors or sSFR, i.e., within the whole galaxy or within a multiple of the stellar half-mass radius {\it vs} within the aperture of the SDSS $\sim 1$ kpc fibers (or consistently corrected measurements).

In Section \ref{subsec:satellites}, we have demonstrated that the amplitude of the galactic conformity signal is enhanced with the inclusion of satellite galaxies in the primary and secondary samples, and that the far-field term is due exclusively to satellite galaxies. We have therefore inferred that different methods for classifying isolated galaxies and separating centrals from satellites might imply different quantitative results. 
This is in agreement with the findings of \cite{Campbell15}, who have more systematically quantified to what extent errors in group finders can affect color-dependent occupation statistics measured directly with galaxy group catalogues. For example, \cite{Kauffmann13} find a conformity effect for quenched galaxies predominately out to a projected 3 Mpc: given our findings from Fig.~\ref{fig:centrals_only_primary}, this could be associated to a central-central correlation alone, but some of the amplitude is likely due to the inclusion of isolated satellites as well.

We have also tested (but not shown) that measuring conformity in Illustris as a function of 2D projected separations instead of 3D real-space distances reduces the amplitude of the signal, and so, for example, we would expect that a projected conformity signal at any distance would require a 3D signal out to an even larger radius.

Relatively to the averaging technique, \citet{Kauffmann13}, for example, stack their galaxies in each primary bin together, thus weighting more heavily galaxies with more satellites. On the other hand, as noted above, in this work we average the red fractions for each galaxy together, thus equally weighting primary galaxies that, when binned in stellar mass, can differ in halo mass by up to a factor of 10. When we stack instead, the primary difference is that the plateau at large radii in the conformity signal truncates earlier. This is interesting, as it suggests that the farthest-field effects are being driven by primary galaxies in low-density environments (as defined by number of neighbors), whereas the nearest-field are being driven by galaxies in high-density environments. \citet{Paranjape15} suggest that only at the large (> 8 Mpc) scales is the difference between 1-halo and 2-halo conformity discernible. If the choice of whether to stack or not affects the amplitude at these scales, then great care must be taken to understand what signal is expected from assembly bias, in addition to the interpretation complications associated to scale-mixing, given that large-radius effects in 3D can affect, at reduced amplitude, 2D conformity on smaller scales.

Furthermore, in this work, we have split our sample into \textit{red} and \textit{blue} populations, rather than using the full sSFR distribution. As commented in Section \ref{sec:conformity}, this choice can per se affect the detection of a primary-mass dependence of the conformity signal. Splitting sSFR by percentiles has the potential to reveal the overall trend in conformity in a single-mass bin -- in particular, whether the trend is continuous or exhibits sharp features --, while using a global color cut highlights any change in red fraction as a function of mass. Both choices are useful, but comparisons need to be done with care. In this work, measuring conformity based on sSFR percentile binning would accentuate possible differences in the overall sSFR distribution between the simulation and observations, potentially obscuring the nonetheless robust connection between halo age conformity and galactic conformity through the color--halo age relation. 

If the simulated and actual one-point sSFR distributions do not match each other, two-point statistics will not match each other either. In Illustris, both the color--magnitude diagram and the sSFR distribution as a function of stellar mass are in reasonable agreement with the observations. However, it has been shown that a significant fraction of red and quenched galaxies is in fact missing in Illustris at the intermediate and large mass range \citep{Vogelsberger14b, Genel14}, making the Illustris color distribution less bimodal than observed.
In fact, when we split the Illustris sample by sSFR percentiles, as \citet{Kauffmann13} do with the \citet{Guo11} semi-analytical model, we also find a weaker conformity signal than the one inferred from SDSS -- effectively by construction given the narrower range of sSFR. Such discrepancy can point either to differences in the way color or sSFR are being measured, or to choices of the subgrid and galaxy-formation models that are adopted in Illustris but that do not fully capture galaxy processes. 
With regard to the actual measurement technique, it would be indeed ideal to be able to directly measure sSFR in the inner cores of Illustris galaxies using mock fibers, rather than to compare some choice for a simulated global value (e.g., the half-mass or twice-half-mass radius SFR values) with the corrected global SDSS SFRs \citep{Brinchmann04}. We have tried this approach, but at $z = 0$, the mock fibers are barely larger than Illustris' resolution of 700 pc, and so the comparison relative to point 5) above is simply not feasible at this time. 
As for the latter point, we are planning to explore in future analyses the emergence of galactic conformity and of the color--age relation, and specifically to address how the actual shape of the {\it galaxy color distribution} at fixed mass affects the amplitude and shape of the galactic conformity signal: this will be possible thanks to upcoming realizations of the Illustris volume with different prescriptions for the underlying galaxy-formation and subgrid models. 

Finally, the discussion proposed thus far serves to emphasize the degree to which methodology impacts the perceived amplitude and shape of galactic conformity. In a recent work, \citet{Kauffmann15} has presented an independent comparison of conformity-like signals in Illustris and SDSS, by making use of the publicly available Illustris data \citep{Nelson15} and by concluding that the simulated conformity effects do not show comparable strength to those seen in observations. It bears emphasizing clearly that our primary results and the conformity signal from Illustris presented by \citet{Kauffmann15} in Fig.~8 are not in conflict, but two different analyses of the same underlying simulation. Whereas \citet{Kauffmann15} presents Illustris data  to directly compare to observations, our focus is on understanding the conformity signal in the Illustris Simulation.
In fact, the apparent qualitative discrepancies are due to observationally-motivated choices as the one discussed above. The difference in sSFR distributions, together with the employment of projected separations and stacking techniques, is bound to suppress the measured conformity signal, therefore explaining why the Illustris-based results in \citet{Kauffmann15} appear weaker than the one presented here.
In turn, the fact that the conformity signal in Fig.~8 of \citet{Kauffmann15} is weaker than the one inferred from SDSS data might indeed imply that in our simulation galactic conformity is underestimated: it is reasonable to speculate that a stronger bimodality in the simulated Illustris galaxy population or simply a different galaxy color distribution that better matches observations could enhance the theoretically-measured conformity signal herein presented.

\section{Summary and Conclusions}
\label{sec:conclusion}

In this paper we have presented a series of phenomenological measurements of the galactic and halo clustering signals and of the relation between galaxy colors and dark-matter halo ages from the redshift $z=0$ snapshot of the Illustris simulation. This is a full-volume cosmological hydrodynamical simulation from which we have selected a sample of about 18,000 galaxies with stellar mass $> 2\times 10^9 \msun$ in a $75 \hmpc$ box. Rather than compare directly with observations, we have chosen to analyze the conformity signal found in Illustris and sought to explain it in the context of dark-matter halo clustering. Further, we have attempted to understand what confounding effects may affect the inferred amplitude and shape of the galactic conformity signal obtained from actual observations.\\

Our primary results are as follows:

\begin{enumerate}

\item We find that the colors of galaxies at fixed stellar or halo mass are spatially correlated for pair separations that go well beyond the virial radius. Namely, we find in Illustris that red isolated galaxies are surrounded by a higher fraction of red neighbors than their bluer counterparts, in qualitative agreement with recent observational claims.\\

\item This galactic conformity signal extends out to $\sim 10$ Mpc for low mass primary galaxies (far-field term), has a larger amplitude at separations smaller than about 3 Mpc (near-field term), and decreases in overall amplitude toward higher primary mass (Fig.~\ref{fig:smconformity}).\\

\item In concordance with previous gravity-only numerical studies, we find a qualitatively similar dark-matter halo conformity signal, in which older haloes are preferentially neighbored by older haloes, at fixed mass (Fig.~\ref{fig:dmonly_conformity}).\\

\item We show that a galaxy color--halo age relation naturally arises in the Illustris Simulation, with the reddest galaxies being preferentially found in the oldest dark-matter haloes (Fig.~\ref{fig:colorage}). \\

\item We demonstrate that a plausible interpretation of the galactic conformity effect can be given as the combination of the halo conformity signal with the relation between galaxy colors and halo ages. \\

\item In practice, we show that by using abundance and age matching, we can reproduce the signal measured directly in the Illustris simulation, demonstrating that once a reasonably tight relationship exists between galaxy property (in this case, color) and the underlying halo property itself clustered (in this case, halo age), then galactic conformity is, to first approximation, a consequence of halo conformity. Conversely, increasing the scatter in the color--halo age relationship lowers the amplitude of the so-modeled conformity, suggesting that strong feedback mechanisms that remove this galaxy-halo correlation may eliminate entirely any galactic conformity signal. In this picture, the underlying dark matter clustering is the primary contributing factor to the galactic conformity signal, rather than a spatial correlation induced by baryonic processes. A color--halo age relation must necessarily be present for a galactic conformity signal to be visible. \\

\item In fact, different proxies for the assembly history of haloes (different halo formation time choices, as well as knowledge of whether a galaxy is a central or satellite) can give substantially different quantitative results, even as the qualitative picture remains the same. Thus, understanding the primary way in which assembly history affects galaxy properties remains an outstanding issue.\\

\item Finally, we show that the conformity signal out to $\sim 3$ Mpc arises from both correlations between central galaxies and from central-satellite correlations. Therefore, there is a 2-halo effect in Illustris accounting for the near-field conformity signal, as well as a 1-halo term. At larger radii, the signal is driven by the correlation between central galaxies and the satellites of other centrals.\\
\end{enumerate}

We note that the actual quantitative amplitude of the theoretically-predicted conformity signal ultimately depends on the simulated color-magnitude diagram (or equivalently, the sSFR vs stellar mass distribution). Given the lack of a strong bimodality of Illustris galaxies colors at intermediate masses and given the lack of a significant fraction of quenched galaxies towards the high-mass end, it is likely that the quantitative galactic conformity results presented in Fig.~\ref{fig:smconformity} underpredict the observational conformity. Nonetheless, this paper shows that \textit{even with} a significantly smaller bimodality, halo assembly bias is an effective mechanism for producing a galactic conformity signal in Illustris.Spatial correlations or interactions between baryons in different haloes are not needed; all that is required is a strong enough correlation between halo age and galaxy color. However, we do not rule out the possibility that the color-halo age relation could be induced by baryonic physics that affect volumes beyond the virial radius. As new measurements are made in different spatial fields and at higher redshifts, the selection criteria and the choice of stacking techniques may have a dramatic effect on the amplitude and shape of the conformity signal inferred from observations, and therefore, on its interpretation.

\section*{Acknowledgments}

ADB, AP, and SG thank Andrew Hearin for helpful discussions and comments on the manuscript and thank Frank van den Bosch, Jeremy Tinker, Guinevere Kauffmann, Alexie Leauthaud, and Cameron McBride for useful discussions. ADB thanks the OSU CCAPP for hosting its 2014 HOD workshop. AP acknowledges support from the HST grant HST-AR-13897. SG acknowledges support provided by NASA through Hubble Fellowship grant HST-HF2-51341.001-A awarded by the STScI, which is operated by the Association of Universities for Research in Astronomy, Inc., for NASA, under contract NAS5-26555. PT acknowledges support from NASA ATP Grant NNX14AH35G. VS acknowledges support through the European Research Council under ERC-StG grant EXAGAL-308037. LH acknowledges support from NASA grant NNX12AC67G and NSF grant AST-1312095.

The Illustris-1simulation was run on the CURIE supercomputer at CEA/France as part of PRACE project RA0844, and the SuperMUC computer at the Leibniz Computing Centre, Germany, as part of GCS-project pr85je. The further simulations were run on the Harvard Odyssey and CfA/ITC clusters, the Ranger and Stampede supercomputers at the Texas Advanced Computing Center through XSEDE, and the Kraken supercomputer at Oak Ridge National Laboratory through XSEDE. The analysis presented in this paper was conducted on the Harvard Odyssey and CfA/ITC clusters.

\appendix

\label{lastpage}

\end{document}